\pdfoutput=1
\documentclass[conference, table]{IEEEtran}
\usepackage[font={small}]{caption}
\usepackage{algpseudocode,algorithm}

\makeatletter
\algrenewcommand\ALG@beginalgorithmic{\normalsize}
\makeatother

\usepackage{comment}
\usepackage{graphicx}
\usepackage{epstopdf}
\usepackage{float}
\usepackage{microtype}
\usepackage{amsmath}
\everypar{\looseness=-1}
\usepackage{booktabs}
\usepackage{color}
\usepackage{longtable}
\usepackage[flushleft]{threeparttable}
\usepackage{rotating}
\usepackage{tabularx}

\def\ie{{i.e.},~}
\def\eg{{e.g.},~}

\usepackage{enumitem}
\usepackage{tabularx}
\usepackage{fancyvrb}
\newcommand{\verbatimfont}[1]{\def\verbatim@font{#1}}%
\usepackage{listings}
\usepackage{tikz}
\usepackage{pifont}
\usepackage{hyperref}

\DeclareRobustCommand*\circled[1]{\tikz[baseline=(char.base)]{ \node[shape=circle,draw,color=white,fill=black,inner sep=0.5pt] (char){#1};}}
\usepackage{xspace}
\newcommand{\iotbench}{{\textsc{\small{IoTBench}}}\xspace}
\newcommand{\SainT}{{\textsc{\small{SainT}}}\xspace}

\usepackage{caption}
\usepackage{subcaption}

\usepackage{multirow}
\usepackage{pgf}
\usepackage{tikz}
\usepackage[utf8]{inputenc}
\usetikzlibrary{arrows,automata}
\usetikzlibrary{matrix,positioning}
\usepackage{bbding}
\tikzset{
    state/.style={
           rectangle,
           rounded corners,
           draw=black, very thick,
           minimum height=2em,
           inner sep=2pt,
           text centered,
           },
}
\usepackage{listings}
\usepackage{color}
\DeclareCaptionFont{black}{\color{black}}
\newenvironment{nstabbing}
  {\setlength{\topsep}{0pt}%
   \setlength{\partopsep}{0pt}%
   \tabbing}
{\endtabbing}

\makeatletter
\def\BState{\State\hskip-\ALG@thistlm}
\makeatother
\ifCLASSINFOpdf
\else
\fi
\usepackage{authblk}
\IEEEoverridecommandlockouts

\begin{document}
\title{\huge{Sensitive Information Tracking in Commodity IoT}}

\setcounter{Maxaffil}{2}
\author[1]{Z. Berkay Celik*\thanks{* contributed equally to this work.}}
\author[2]{Leonardo Babun*}
\author[2]{Amit K. Sikder}
\author[2]{Hidayet Aksu}
\author[1]{\\Gang Tan}
\author[1]{Patrick McDaniel}
\author[2]{A. Selcuk Uluagac\vspace{-6pt}}

\affil[1]{Department of CSE, The Pennsylvania State University\authorcr
{\texttt{\{zbc102,gtan,mcdaniel\}@cse.psu.edu}}}
\affil[2]{Department of ECE, Florida International University\authorcr
{\texttt{\{lbabu002,asikd003,haksu,suluagac\}@fiu.edu}}}

\newcommand{\itembase}[1]{\setlength{\itemsep}{#1}}
\newcommand{\showextended}{1}% set to 0 to hide problems
\newcommand{\extended}[1]{\ifthenelse{\equal{\showextended}{1}}{#1}{}}

\maketitle

\begin{abstract}
Broadly defined as the Internet of Things (IoT), the growth of commodity devices that integrate physical processes with digital connectivity has had profound effects on society--smart homes, personal monitoring devices, enhanced manufacturing and other IoT apps have changed the way we live, play, and work. Yet extant IoT platforms provide few means of evaluating the use (and potential avenues for misuse) of sensitive information. Thus, consumers and organizations have little information to assess the security and privacy risks these devices present. In this paper, we present \SainT, a static taint analysis tool for IoT applications. \SainT operates in three phases; (a) translation of platform-specific IoT source code into an intermediate representation (IR), (b) identifying sensitive sources and sinks, and (c) performing static analysis to identify sensitive data flows. We evaluate \SainT on 230 SmartThings market apps and find 138 (60\%) include sensitive data flows. In addition, we demonstrate \SainT on \iotbench, a novel open-source test suite containing 19 apps with 27 unique data leaks. Through this effort, we introduce a rigorously grounded framework for evaluating the use of sensitive information in IoT apps---and therein provide developers, markets, and consumers a means of identifying potential threats to security and privacy.
\end{abstract}
\begin{IEEEkeywords}
IoT privacy, static taint analysis
\end{IEEEkeywords}

\section{Introduction}
\label{sec:intro}
The introduction of IoT devices into public and private spaces has changed the way we live.  For example, home applications supporting smart locks, smart thermostats, smart switches, smart surveillance systems, and Internet-connected appliances change the way we monitor and interact with our living spaces.  Here mobile phones become movable control panels for managing the environment that supports entertainment, cooking, and even sleeping.  Such devices enable our living space to be more autonomous, adaptive, efficient, and convenient.  However, IoT has also raised concerns about the privacy of these digitally augmented spaces~\cite{ronen2017iot, fernandes2016security, jia2017contexiot, ho2016smart}. These networked devices have access to data that can be intensely private, e.g., when you sleep, what your door lock pin code is, what you watch on TV or other media, and who and when others are in the house.  Moreover, the state of the devices themselves represents potentially sensitive information.

Because IoT apps are exposed to a myriad of sensitive data from sensors and devices connected to the hub, one of the chief criticisms of modern IoT systems is that the existing commercial frameworks lack basic tools and services for analyzing what they do with that information--i.e., application privacy~\cite{zeng2017end, naeini2017privacy, yang2017survey}. SmartThings~\cite{samsung}, OpenHab~\cite{openhab}, Apple's Homekit~\cite{apple} provide guidelines and policies for regulating security~\cite{Official, OpenHabGuideline, appleSecurity}, and related markets provide a degree of internal (hand) vetting of the applications prior to distribution~\cite{smartThings-review, AppleHomekitReview}. However, tools for evaluating privacy risks in IoT implementations is at this time largely non-existent. What is needed is a suite of analysis tools and techniques targeted to IoT platforms that can identify privacy concerns in IoT apps. This work seeks to explore formally grounded methods and tools for characterizing the use of sensitive data, and identifying the sensitive data flows in IoT implementations.

Current sensitive data tracking tools designed for mobile apps and other domains~\cite{EnckTaintDroid, ArztFlowdroid} have proved to be inadequate for several reasons~\cite{fernandes2016flowfence, jia2017contexiot}. First, current tools may miss sources (\eg sensor state (locked/unlocked)) and sinks (\eg a network connection) designed for IoT; thus, they can be circumvented by malicious apps with ease. Second, security-critical design flaws in the permission model of IoT platforms, for instance, over-privilege device controls due to the current coarse-grained access controls~\cite{fernandes2016security}, requires the analysis responsive to these permissions and their effects. Lastly, IoT-specific implementations such as state variables and web service IoT apps largely differs from other platforms~\cite{smartThings-documentation}; therefore, on-demand algorithms are required to maintain precision.

In this paper, we present \SainT, a static taint analysis tool for IoT apps. \SainT finds sensitive data flows in IoT apps by tracking information flow from sensitive sources, \eg device state (locked/unlocked) and user info (away/at home) to external sinks, \eg Internet connections, and SMS.  We conduct a study of three major existing IoT platforms (\ie SmartThings, OpenHAB, and Apple's HomeKiT) to identify IoT-specific sources and sinks as well as their sensor-computation-actuator program structures. We then translate source code of an IoT app into an intermediate representation (IR). The \SainT IR models the app's lifecycle--including program entry points, user inputs, sensor states--as well simplifying analysis by abstracting away code that is not relevant to information flow. In this, we address IoT-specific challenges like events/actions and asynchronously executing events, as well as platform-specific challenges such as call by reflection and the use of state variables. Thereafter \SainT uses the IR to perform efficient static analysis that tracks information flow from sensitive sources to sink outputs. 

We present two studies validating and demonstrating \SainT. The first is a horizontal market study in which we evaluated 230 Smarthings IoT apps including 168 market vetted (called official) and 62 non-vetted (called third-party) apps. \SainT correctly flagged 92 out of 168 official and 46 out of 62 third-party apps exposing at least one sensitive data via the Internet or messaging services. Further, the study showed that half of the analyzed apps transmit out at least three different sensitive data sources (e.g., device info, device state, user input) via messaging or Internet. Similarly, approximately two-thirds of the apps define at most two separate sensitive sink interfaces and recipients (e.g., remote hostname or URL for Internet and contact information for messaging). In a second study, we introduced \iotbench, an open-source application corpus for validating IoT analysis. Our analysis of \SainT on \iotbench showed that it correctly identified the 25 of the 27 unique leaks in the 19 apps.  \SainT produced two false-positives that were caused by flow over-approximation resulting from reflective methods calls. Additionally, the two missed code sites contained side-channel leaks and therefore were outside the scope of \SainT analysis. 

It is important to note that the code analysis identifies {\bf potential flows of sensitive data}.  What the user does with a discovered sensitive data flow is outside the scope of \SainT. 
Indeed, the importance of a flow is highly contextual--one cannot divine the impact or correctness of a flow without understanding the environment in which it is deployed--whether the exposure of a camera image, the room temperature, or television channel represents a privacy concern depends entirely on who and under what circumstances the device and app is used.
Hence, we identify those flows which have the potential impact on user or environmental security and privacy.  We expect that the results will be recorded and the code hand-investigated to determine the cause(s) of the data flows. If the data flow is deemed malicious or dangerous for the domain or environment, the app can be rejected (from the market) or modified (by the developer) as needs dictate. 

\vspace{3pt}

\noindent
We make the following contributions:

\begin{itemize}[noitemsep,nolistsep]
\item We introduce the \SainT system that automates information-flow tracking using inter- and intra-data flow analysis on an IoT app.

\item We evaluate \SainT on 230 IoT apps and expose sensitive information use in commodity apps.

\item We validate \SainT on a new open-source IoT-specific test corpus \iotbench, an open-source repository of 19 malicious hand-crafted apps.

\end{itemize}

\noindent We begin in the next section by defining the analysis task and outlining the security and attacker models.

\section{Problem Scope and Attacker Model}
\label{sec:background}
\noindent\textbf{Problem Scope.} \SainT analyzes the source code of an IoT app, identifies sensitive data from a \emph{taint source}, and attaches taint labels that describe sensitive data's sources and types. It then performs static taint analysis that tracks how labeled data (source data, e.g., camera image) propagates in the app (sink, e.g., network interface). Finally, it reports cases, where sensitive data transmits out of the app at a \emph{taint sink} such as through the Internet or some messaging service. In a warning, \SainT reports the source in the taint label and the details about the sink, such as the external URL or the mobile phone number. \SainT does not determine whether the data leaks are malicious or dangerous; however, the output of \SainT can be further analyzed to verify whether an app conforms its functionality and notify users to make informed decisions about potential privacy risks, e.g., camera image is transmitted.

Currently, \SainT is designed to analyze SmartThings IoT apps written in the Groovy programming language. We evaluate the SmartThings platform for two reasons. First, it supports the largest number of devices (142) among all IoT platforms and provides apps of various functionalities~\cite{smartThings-devices}. Second, it has a detailed publicly available documentation that helps validate our findings~\cite{smartThings-documentation}. As we will detail in Section~\ref{sec:iotlife}, \SainT exploits the highly-structured nature of the IoT programming platforms and extracts an abstract intermediate representation from the source code of an IoT app. This would allow the algorithms developed in \SainT to be effectively integrated into other programming platforms written in different programming or domain-specific languages.

\noindent\textbf{Attacker Model.} \SainT detects sensitive data flows from taint sources to taint sinks caused by carelessness or malicious intent. We consider an attacker who provides a user a malicious app that is used to leak sensitive information with or without permissions granted by the user. First, the granted permissions may violate user privacy by deviating from the functionality claimed by the app. Second, permissions granted by an IoT programming platform may also be used to leak information; for instance,  permissions to access the hub id or the manufacturer name are often granted by default to develop device-specific solutions. We assume attackers cannot bypass the security measures of an IoT platform, nor can they exploit side channels~\cite{amit}. For instance, an app that changes the light intensity to leak the information about whether nobody is at home is out of the scope of this work.

\section{Background of IoT Platforms}
\label{sec:prog-frameworks}
We present background of the SmartThings IoT platform~\cite{smartThings-documentation} to gain insights into the structure of its apps. We also investigate two other popular IoT platforms: OpenHab~\cite{openhab} and Apple's Homekit~\cite{apple}. We find that they use similar programming structures and the differences lie in only the communication protocols between IoT devices and edge systems. Lastly, we define each potential type of taint sources, the mechanisms for taint propagation, and taint sinks by studying their API documentation. 

\subsection{Overview of IoT Platforms}
\noindent\textbf{SmartThings} is a proprietary platform developed by Samsung. The platform includes three components: a hub, apps, and the cloud backend~\cite{smartThings-review}. The hub controls the communication between connected devices, the cloud backend, and mobile apps. Apps are developed with Groovy (a dynamic, object-oriented language) in a Kohsuke sandboxed environment~\cite{fernandes2016security}. The sandbox limits developers to a specific subset of the Groovy language for performance and security. For instance, the sandbox bans apps from creating their own classes and threads. The cloud backend creates software wrappers for physical devices and runs the apps. 

The permission system in SmartThings allows a developer to specify devices and user inputs required for an app at install time. User inputs are used to implement the app logic. For instance, a user input is used to set the heating point of a thermostat. Devices in SmartThings have capabilities (\ie permissions). Capabilities are composed of \emph{actions} and \emph{events}. Actions represent how to control or actuate devices and events represent the state information of devices. Actions and events are not one to one. While a device may support many events, it may have limited actions. Apps are event-driven. They subscribe to device events or other pre-defined events such as clicking an icon; when an event is activated, the corresponding event handler is invoked to take actions.

Users can install SmartThings apps in two different ways using a smartphone companion app called SmartThings Mobile. First, users may download apps through the official app market. Second, users may install third-party apps through the Web IDE on a proprietary cloud backend. Publishing an app in the official market requires the developer to submit the source code of the app for review. Official apps appear in the market after the completion of a review process that takes around two months to finish~\cite{smartThings-review}. Users can also develop or install the source code of a third-party app and make it accessible to only themselves using the Web IDE. These apps do not require any review process and are often shared in the SmartThings community forum~\cite{Community}. Compared to other competing platforms, Smartthings supports more devices and has a growing number of official and third-party apps.

\noindent\textbf{OpenHAB} is a vendor- and technology-agnostic open-source automation platform built in the Eclipse IDE~\cite{openhab}. It includes various devices specifically designed for home automation. OpenHAB is open sourced and provides flexible and customizable device integration and applications (referred to as rules) to build automated tasks. Similar to the SmartThings platform,  the rules are implemented through three triggers to react to the changes in the environment. Event-based triggers listen to commands from devices; timing-based triggers respond to special times (\eg midnight); system-based triggers run with certain system events such as system start and shutdown. The rules are written in a Domain Specific Language (DSL) based on the Xbase language, which is similar to the Xtend language with some missing features~\cite{efftinge2012xbase}. Users can install OpenHAB apps by placing them in rules folder of their installations and from Eclipse IoT Marketplace~\cite{openHABMarket}.

\noindent\textbf{Apple's HomeKit} is a development kit that manages and controls compatible smart devices~\cite{apple}. The interaction between users and devices occurs through Siri and HomeKit apps. Similar to SmartThing and OpenHAB, each device has capabilities that represent what a device can do. Actions are defined to send commands to specific devices and triggers can be defined to execute actions based on location, device, and time events. Developers write scripts to specify a set of actions, triggers, and optional conditions to control HomeKit-compatible devices. Developing applications in HomeKit can either be written in Swift or Objective C. Users can install HomeKit apps using the Home mobile app provided by Apple~\cite{appleMarket}.

\subsection{IoT Application Structure} 
\label{sec:privacy-hooks}
\begin{figure}[t!]
\begin{center}
\includegraphics[width=1\columnwidth]{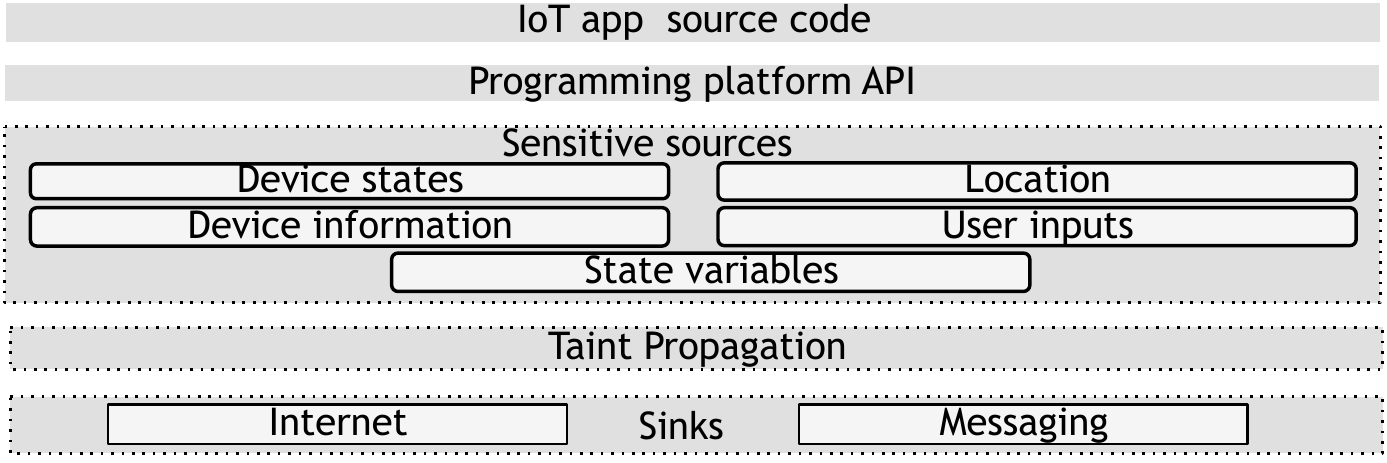}
\caption{\SainT's source and sink categorization in IoT apps.}
\label{fig:categories}
\end{center}
\end{figure}

From our studying of the three IoT platforms, we found that their apps share a common structure and common types of taint sources and sinks. In this section, we describe these common taint sources and taint sinks to understand why they pose privacy risks and how sensitive information gets propagated in their app structure (see Fig.~\ref{fig:categories}). We give the specific list of SmartThings APIs, by far the most comprehensive one available, that access taint sources and taint sinks in Appendix~\ref{appendix:sources-sinks}. We will use this list in Sec.~\ref{sec:static-taint-tracking} while performing data flow analysis.

\noindent\textbf{Taint Sources.} We classify taint sources into five groups based on information types. 

\noindent\textbf{1) Device States.} Device states are the attributes of a device. An IoT app can acquire a variety of privacy-sensitive information through device state interfaces. For instance, a door-lock interface returns the status of the door as locked or unlocked. In our analysis, we marked device states sensitive as they can be used to profile habits of a user and pose risks to physical privacy.

\noindent\textbf{2) Device Information.} IoT apps grants access to IoT devices at install time. Our investigations reveal the platforms often define interfaces to access device information such as its manufacturer name, id, and model. This allows a developer to write device-specific apps. We mark all interfaces used to acquire device information as sensitive as they can be used for marketing and advertisement. Note that device information is static and does not change over the course of app execution. In contrast, device states introduced earlier may change during app execution; for instance, an action of an app may change a device's state.

\noindent\textbf{3) Location.} In the IoT domain, location information refers to a user's geolocation or geographical location. Geolocation defines a virtual property such as a garage or an office defined by a user to control devices in that location. Geographical location is used to control app logic through time zones, longitudes, and latitudes. This information is often provided by the programming platform using the ZIP code of the user at install time. For instance, local sunrise and sunset times of a user's location may be used to control the window shade of a house. Location information is acquired through location interfaces; therefore, we mark these interfaces as taint sources.

\noindent\textbf{4) User Inputs.} IoT apps often require user inputs either to manage app logic or to control devices. In a simple example, a temperature value needs to be entered by a user at install time to set the heating point of a thermostat. User inputs are also often used to form predicates that control device actions; for instance, an app may turn off the switch of a device at a particular time entered by the user. Lastly, users may enter contact information to enable notifications through messaging services when specific events occur. We mark such inputs as sensitive since they contain personally identifiable data and may be used to profile user behavior. We will discuss more about the semantics of user inputs in Section~\ref{sec:limit-discuss}. 

\noindent\textbf{5) State Variables.} IoT apps do not store data about their previous executions. To retrieve data across executions, platforms allows apps to persist data to some propriety external storage and retrieve this data in later executions. For instance, an app may persist a ``counter'' that keeps track of how many times a door is unlocked; during every execution of the app, the counter is retrieved from external storage and incremented when a door is unlocked. We call such persistent data app \emph{state variables}. As we detail in Sec. ~\ref{sec:advanced-taint-tracking}, state variables store sensitive data, and needs to be tracked during taint propagation.  

\noindent\textbf{Taint Propagation.} An IoT app invokes actions to control its devices when a particular event occurs. Actions are invoked in event handlers and may change the state of the devices. For instance, when a motion sensor triggers a sensor-active event, an app may invoke an event handler to take an action that changes the state of the light switch from off to on. This is a straightforward approach to invoke an action. Event handlers are not limited to implement only device actions. Apps often call other functions for implementing the app logic, sending messages, and logging device events to an external database. 

During the execution of event handlers, it is necessary to track how sensitive information propagates in an app's logic. To obtain precision in taint propagation, we start from event handlers to propagate taint when tainted data is copied or used in computation, and we delete taint when all traces of tainted data are removed (\eg when some variable is loaded with a constant). We will detail event handlers and \SainT's taint propagation logic in Sec.~\ref{sec:methodology}.

\noindent\textbf{Taint Sinks.} Our initial analysis also uses two taint sinks (although adding more later is a straightforward exercise).

\noindent\textbf{1) Internet.} IoT apps may send sensitive data to external services or may act as web services through which external entities acquire sensitive information. For the first kind, HTTP interfaces may be used to send out information. For instance, an app may connect to a weather forecasting service (\eg www.weather.com) and send out its location information to get the local weather. For the second kind, a web-service IoT app may expose a URL that allows external entities to make requests to the app. For instance, a request from a remote server may be used to get the room temperature value. We will detail how \SainT tracks taint of web services apps in Sec.~\ref{sec:advanced-taint-tracking}. 

\noindent\textbf{2) Messaging Services.} IoT apps use messaging APIs to deliver push notification to mobile-app users and to send SMS messages to designated recipients when specific events occur. We consider all messaging service interfaces taint sinks--naturally, as they exfiltrate data by design.

\begin{figure}[!t]
    \centering{\includegraphics[width=1\columnwidth]{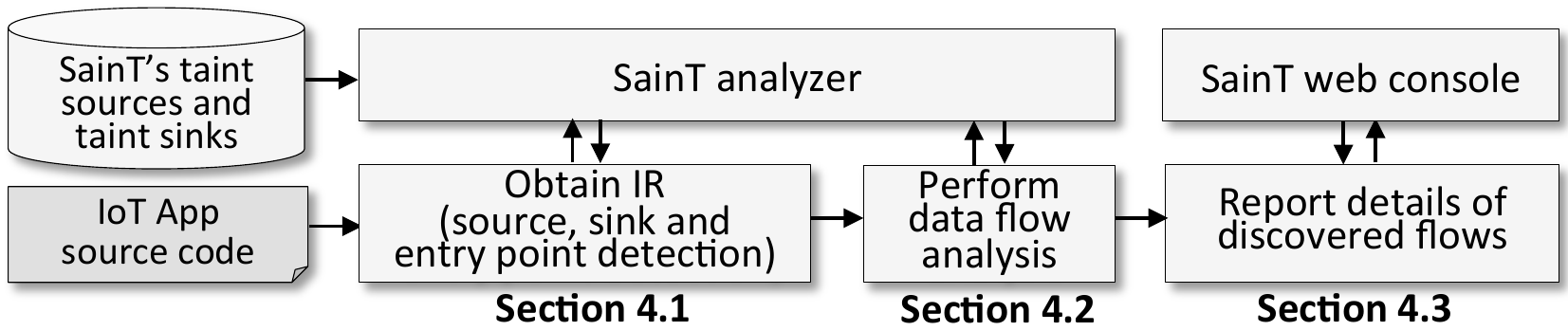}}
    \caption{Overview of \SainT architecture.}
    \label{fig:Arch}
\end{figure}

\section{\textsc{SainT}}
\label{sec:methodology}
We present \SainT, a static taint analysis tool designed and implemented for SmartThings apps. Fig.~\ref{fig:Arch} shows the overview of \SainT architecture. We implement the \SainT analyzer that extracts an intermediate representation (IR) from the source code of an IoT app. The IR is used to construct an app's entry points, event handlers, and call graphs (Sec.~\ref{sec:iotlife}). Using these, \SainT models the lifecycle of an app and performs static taint analysis (Sec.~\ref{sec:static-taint-tracking}). Finally, based on static taint analysis, it reports sensitive data flows from sources to sinks; for each data flow, the type of the sensitive information, as well as information about sinks, are reported (Sec.~\ref{sec:saint_implementation}). 
 
\subsection{From Source Code to IR}
\label{sec:iotlife}
The first step toward modeling the app lifecycle is to extract an IR from an app's source code. We exploit the highly-structured nature of IoT programming platforms based on our analysis in Sec. \ref{sec:prog-frameworks}. We found that IoT systems are generally structured similarly regardless of their purpose and complexity. The dominant IoT systems structure their app's design around the \emph{sensor-computation-actuator} idioms. Therefore, we translate the source code of an IoT app into an IR by exploiting this structure.  

\begin{figure}[t!]
\begin{center}
\includegraphics[width=1\columnwidth]{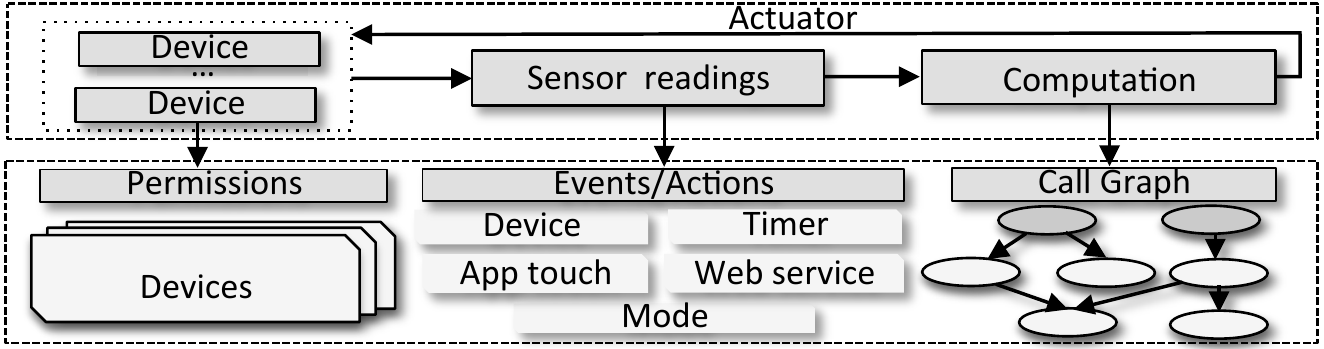}
\caption{Components of the Intermediate Representation (IR).}
\label{fig:components}
\end{center}
\end{figure}

\SainT builds the IR from a framework-agnostic component model, which is comprised of the building blocks of IoT apps, shown in Fig.~\ref{fig:components}. A broad investigation of existing IoT environments showed that the programming environments could be generalized into three component types: (1) \textit{Permissions} grant capabilities to devices used in an app; (2) \textit{Events/Actions} reflect the association between events and actions (when an event is triggered, an associated action is performed); and (3) \textit{Call graphs} represent the relationship between entry points and functions in an app. The IR has several benefits. First, it allows us to precisely model the app lifecycle as described above. Second, it is used to abstract away parts of the code that are not relevant to property analysis, \eg \texttt{definition} blocks that specify app meta-data and \texttt{logger} logging code. Third, it allows us to have effective taint tracking, \eg by associating permissions with the corresponding taint tags and by knowing what methods are entry points. 

\begin{figure}[t!]
\begin{center}
\includegraphics[width=0.95\columnwidth]{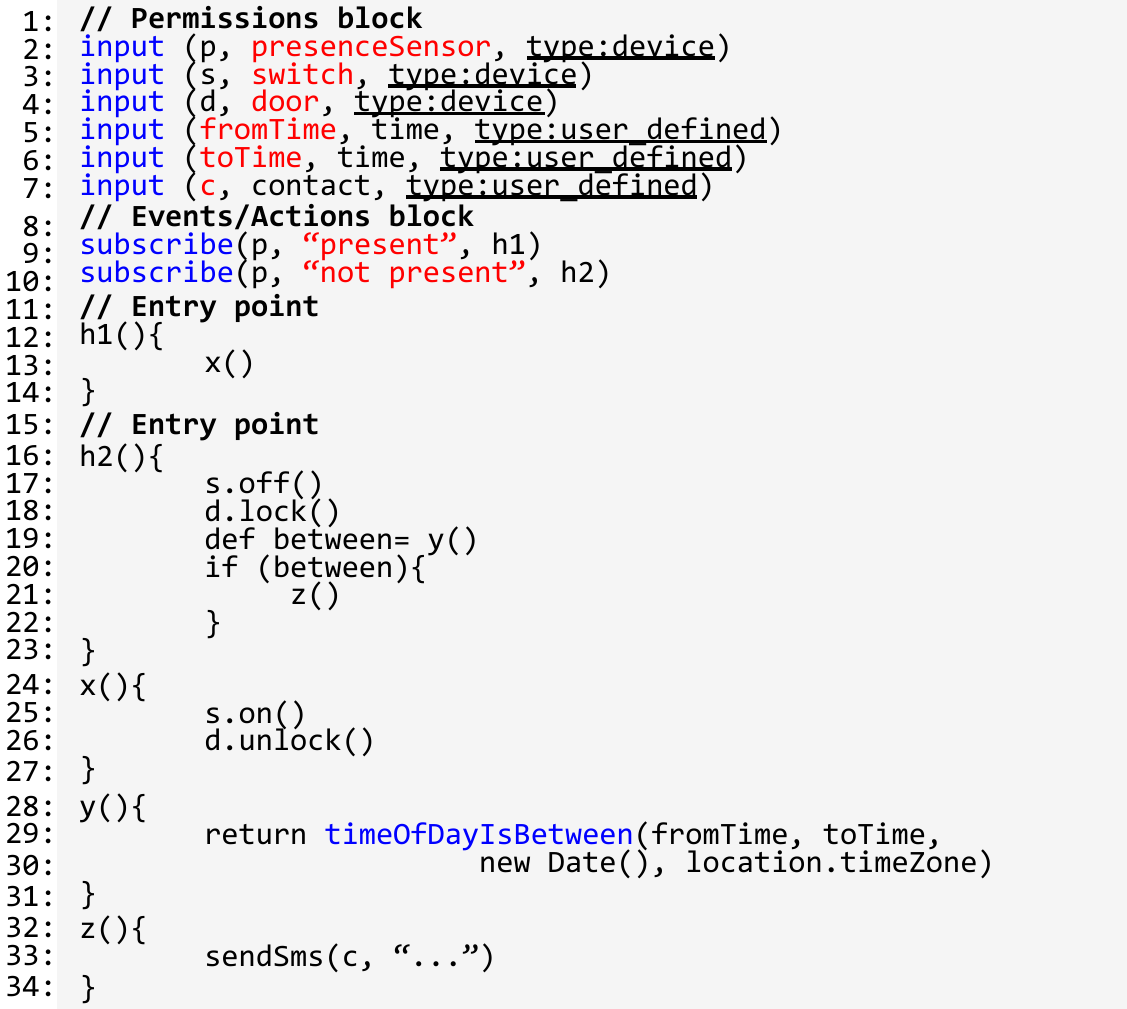}
\caption{The IR of a sample app constructed with \SainT from its source code to demonstrate the precise modelling of an IoT app lifecycle. (Appendix~\ref{appendix:example-app} presents its Groovy source code.)}
\label{fig:motExample}
\end{center}
\end{figure}

Presented in Fig.~\ref{fig:motExample}, we use a sample app to illustrate the use of the IR. The app unlocks the front door and turns on the lights when she arrives at home. When she leaves, it turns off the lights, locks the front door, and sends to the security service a short message that she is away based on the preferred time window specified by her.

\noindent\textbf{Permissions.} Permissions are granted when a user installs or updates an app. This is where various types of devices and user inputs are described and granted access. The permissions are read-only, and app logic is implemented using the permissions. The \SainT analyzer analyzes the source code of an app and extracts permissions for all devices and user inputs. Turning to the IR example in Fig.~\ref{fig:motExample}, the permission block (lines 1-7) defines: (1) the devices: a presence sensor, a switch, and a door; and (2) user inputs: security-service ``contact'' information for sending notification messages, and ``fromTime'' and ``toTime'' values that are used to determine whether notification messages should be sent. For each permission, the IR declares a triple following keyword ``input''. For devices, the first two entries map device identifiers to their platform-specific device names in order to determine the interfaces that a device may access. For instance, an app granting access to a switch may use \texttt{theswitcState} object to access its ``on'' or ``off'' state. For a user input, the line in the IR contains the string name storing the user input and its type. The next entry labels the input with a taint tag showing the type of information such as the user-defined tag. We note that we consider user inputs sensitive. 

We also include in the permission block a set of common interfaces designed for all apps that may leak sensitive data. For instance, \texttt{location.currentMode} gives the location mode either set to home or away. We assign each sensitive value to its label based on taint tags defined in Sec.~\ref{sec:privacy-hooks}. In this way, we obtain a complete list of sensitive interfaces an app may access.

\noindent\textbf{Events/Actions.} Similar to mobile applications, an IoT app does not have a main method due to its event-driven nature. Apps implicitly define entry points by subscribing events. The events/actions block in an IR is built by analyzing how an app subscribes to events. Each line in the block includes three pieces of information: the mapping used for a device, a device event to be subscribed, and an event handler method to be invoked when that event occurs. The event handler methods are commonly used to take device actions. Therefore, an app may define multiple entry points by subscribing multiple events of a device or devices. Turning to our example, the event of state changing to ``present" is associated with an event handler method named \texttt{h1()} and the event of changing to ``not present'' with the \texttt{h2()} method.  

We also found that events are not limited to device events, and can be generated in many other ways: (1) \emph{Timer events}; event handlers are scheduled to take actions within a particular time or at pre-defined times (\eg an event handler is invoked to take actions after a given number of minutes has elapsed or at specific times such as sunset); (2) \emph{Web service events}; IoT programming platforms may allow an app to be accessible over the web. This allows external entities (\eg If This Then That (IFTTT)~\cite{ifttt}) to make requests to the app, and get information about or control end devices; (3) \emph{App touch events}; for example, some action can be performed when the user clicks on a button in an app; (4) what actions get generated may also depend on \emph{mode events}, which are behavior filters that automate device actions. For instance, an app running in ``home" mode turns off the alarm and turns on the alarm when it is in the ``away" mode. Our \SainT analyzer analyzes all event subscriptions and finds their corresponding event handler methods; it creates a dummy main method for each entry point.

\noindent\textbf{Asynchronously Executing Events.} While each event corresponds to a unique event handler, the sequence of the event handlers cannot be decided in advance when multiple events happen at the same time. For instance, in our example, there could be a third subscription in the event/actions block that subscribes to the switch-off event to invoke another event-handler method. We consider eventually consistent events, which means any time an event handler is invoked, it will finish execution before another event is handled, and the events are handled in the order they are received by an edge device (\eg a hub). We base our implementation on path-sensitive analysis that analyzes an app's event handlers, which can run in arbitrary sequential order. This is enabled by constructing a separate call graph for each entry point.

\noindent\textbf{Call Graphs.} We create a call graph for each entry point that defines an event-handler method. Turning to IR depicted in Fig.~\ref{fig:motExample}, we have two entry points \texttt{h1()} and  \texttt{h2()} (line 12 and 16). \texttt{h1()} invokes \texttt{x()} to unlock the door and turn on the lights. Entry point \texttt{h2()} turns off the light and locks the door. It then calls method \texttt{y()} to check the time to decide whether to send a short message to a predefined contact via method \texttt{z()}. We note that the next section will detail how to construct call graphs, for example, in the case of call by reflection.

\subsection{Static Taint Tracking}
\label{sec:static-taint-tracking}
We start with backward taint tracking (Sec.~\ref{sec:basic-taint-tracking}). We then present algorithms to address platform- and language-specific taint-tracking challenges like state variables, call by reflection, web-service IoT apps, and Groovy-specific properties (Sec.~\ref{sec:advanced-taint-tracking}). Last, we discuss the problem of implicit flows in static taint tracking (Sec.~\ref{sec:implicit-flows}). 

\subsubsection{Backward Taint Tracking}
\label{sec:basic-taint-tracking}
From the inter-procedural control flow graph (ICFG) of an app, \SainT's backward taint tracking consists of two steps: (1) it first performs taint tracking backward from taint sinks to construct possible data-leak paths from sources to sinks; (2) using path- and context-sensitivity, it then prunes infeasible paths to construct a set of \emph{feasible paths}, which are the output of \SainT's static taint tracking.

In the first step, \SainT starts at the sinks of the ICFG and propagates taint backward. The reason that \SainT uses the backward approach is to reduce the processing overhead by starting from a few sinks instead of from a huge number of sensitive sources. This is confirmed by checking the ratio of sinks over sources in analyzed IoT apps (see Fig.~\ref{fig:sensitive_source_leaks} in Sec.~\ref{sec:evaluation} for taint source analysis and see Fig.~\ref{fig:sink-analysis} in Sec.~\ref{sec:evaluation} for taint sink analysis). 

\newcommand*\Let[2]{\State #1 $\gets$ #2}
\algnewcommand\algorithmicinput{\textbf{Input:}}
\algnewcommand\INPUT{\item[\algorithmicinput]}
\algnewcommand\algorithmicoutput{\textbf{Output:}}
\algnewcommand\OUTPUT{\item[\algorithmicoutput]}

\newcommand*{\worklist}{\mathit{worklist}}
\newcommand*{\done}{\mathit{done}}
\newcommand*{\taintpath}{\mathit{path}}
\newcommand*{\depend}{\mathit{dep}}
\newcommand{\predSet}[2]{\{#1\,\mid\,#2\}}
\newcommand{\ident}{\mathit{id}}
\newcommand{\idents}{\mathit{ids}}

\begin{algorithm}[t]
\footnotesize
%\setstretch{0.9}
\begin{algorithmic}[1]
\INPUT{$ICFG:$ Inter-procedural control flow graph}
\OUTPUT{Dependence relation $\depend$}
%\State
\vspace{1pt}
\State $\worklist \gets \emptyset;\ \done \gets \emptyset;\ \depend \gets \emptyset$
\For{an $\ident$ in a sink call's arguments at node $n$}
    \Let{$\worklist$}{$\worklist \cup \{(n,\ident)\}$}
\EndFor
\While{$\worklist$ is not empty}
  \Let{$(n,\ident)$}{$\worklist.\mathit{pop}()$}
  \Let{$\done$}{$\done \cup \{(n,\ident)\}$}
    \For{node $n'$\hspace*{1mm}with $\ident$ def.\footnotemark{}\hspace*{0.5mm}in\hspace*{0.5mm}assignment\hspace*{0.5mm}$\ident=e$}
        \Let{$\idents$}{$\predSet{(n',\ident')}{\ident'\ \mbox{is an identifier in}\ e}$}
        \Let{$\worklist$}{$\worklist \cup (\idents\ \setminus \ \done)$}
        \Let{$\depend$}{$\depend \cup \{(n:\ident, n':\idents) \}$}
    \EndFor
\EndWhile
\caption{Computing dependence from taint sinks}
\label{algo:varTainting}
\end{algorithmic}
{\small{{$^{1}$ An $\ident$ definition  means that there is a control-flow path from $n'$ to $n$ and on the path there is no other assignments to $\ident$.}}}
\end{algorithm}

Algorithm~\ref{algo:varTainting} details the steps for computing a \emph{dependence} relation that captures how values propagate in an app. It is a worklist-based algorithm. The worklist is initialized with identifiers that are used in the arguments of sink calls. Note that each identifier is also labeled with the node information to uniquely identify the use of an identifier because the same identifier can be used in multiple locations. The algorithm then takes an entry $(n,\ident)$ from the worklist and finds a definition for $\ident$ on the ICFG; it adds identifiers on the right-hand side of the definition to the worklist; furthermore, the dependence between $\ident$ and the right-hand side identifiers are recorded in $\depend$. Note that for ease of presentation the algorithm treats parameter passing in a function call as inter-procedural definitions.

\begin{figure}[!t]
	\centering{\includegraphics[width=\columnwidth]{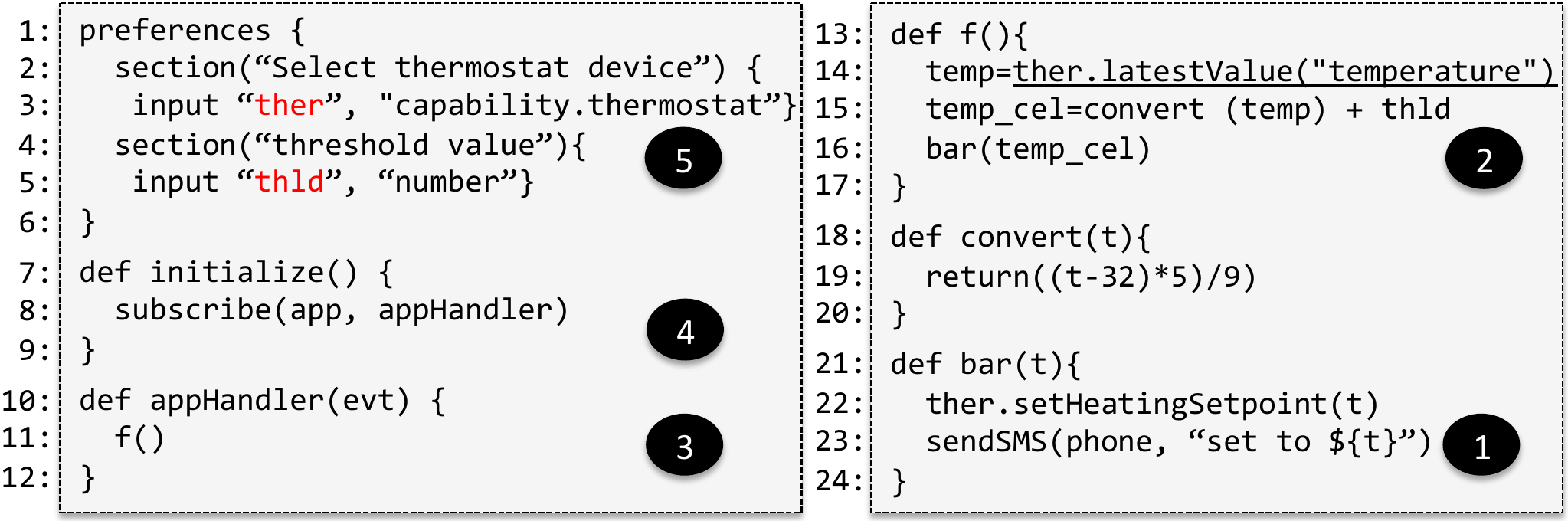}}
	\caption{Taint tracking under backward flow analysis.}
	\label{fig:taintTracking}
\end{figure}

To illustrate, we use the code in Fig.~\ref{fig:taintTracking} as an example. There is a sink call at place \circled{1}. So the worklist is initialized to be ((23:{\tt phone}), (23:{\tt t})); for illustration, we use line numbers instead of node information to label identifiers. Then, because of the function call at \circled{2}, (16:{\tt temp\_cel}) is added to the worklist and the dependence (23:{\tt t}, 16:[{\tt temp\_cel}]) is recorded in $\depend$. With similar computation, the final output dependence relation for the example is as follows:

\vspace{1pt}{\footnotesize{
\begin{nstabbing}
  (23:{\tt t}, 16:[{\tt temp\_cel}]), (16:{\tt temp\_cel}, 15:[{\tt temp}, {\tt thld}]), \\
  (15:{\tt temp}, 14:[{\tt ther.latestValue}])
\end{nstabbing}}}

\noindent With the dependence relation computed and information about taint sources, \SainT can easily construct a set of possible data-leak paths from sources to sinks. For the example, since the threshold value \texttt{thld} is a user-input value (lines 4 and 5 in Fig.~\ref{fig:taintTracking}), we get the following possible data-leak path: 5:\texttt{thld} to 16:{\tt temp\_cel} to 23:{\tt t}.

In the next step, \SainT prunes infeasible data-leak paths using path- and context-sensitivity. For a path, it collects the evaluation results of the predicates at conditional branches and checks whether the conjunction of those predicates (i.e., the path condition) is always false; if so, the path is infeasible and discarded.\footnote{Similar to how symbolic execution prunes paths via path conditions.} For instance, if a path goes through two conditional branches and the first branch evaluates $x>1$ to true and the second evaluates $x<0$ to true, then it is an infeasible path. \SainT does not use a general SMT solver to check path conditions. We found that the predicates used in IoT apps are extremely simple in the form of comparisons between variables and constants (such as $x=c$ and $x>c$); thus, \SainT implemented its simple custom checker for path conditions. Furthermore, \SainT throws away paths that do not match function calls and returns (using depth-one call-site sensitivity). At the end of the pruning process, we get a set of feasible paths from taint sources to sinks.

\subsubsection{SmartThings Idiosyncrasies} 
\label{sec:advanced-taint-tracking}
Our initial prototype implementation of \SainT was based on the taint tracking approach we discussed. However, SmartThings platform has a number of idiosyncrasies that may cause imprecision in taint tracking. We next discuss how these issues are addressed in \SainT.

\noindent\textbf{Field-sensitive Taint Tracking of State Variables.} As discussed before, IoT apps use state variables that are stored in external storage to persist data across executions. In SmartThings, state variables are stored in either the global \texttt{state} object or the global \texttt{atomicState} object. Listing~\ref{listing-general} (lines 1--9) presents an example app using the \texttt{state} object to store a field named \texttt{switchCounter} to track the number of times a switch is turned on. To taint track potential data leaks through state variables, \SainT applies field-sensitive analysis to track the data dependencies of all fields defined in the \texttt{state} and \texttt{atomicState} objects. We label fields in those two objects with a new taint label ``state variable" and perform taint tracking. For instance,  the \texttt{taintedVar} variable in Listing~\ref{listing-general} is labeled with the state-variable taint by \SainT.

\begin{lstlisting}[float=t!, caption= Sample code blocks for SmartThings idiosyncrasies ,label=listing-general]
/*  A code block of an app using a state variable */ 
def initialize() {
    state.switchCounter = 0
    subscribe(theswitch, "switch.on", turnedOnHandler)
}
def turnedOnHandler() {
    state.switchCounter = state.switchCounter + 1
    taintedVar = state.switchCounter // tainted
}
/* A code block of app using call by reflection */
def getMethod(){
  httpGet("http://url"){ 
    resp -> if(resp.status == 200){
                methodName = resp.data.toString()
            }
    "$methodName"() //call by reflection
}           
def foo() {...}
def bar() {...}
/* A code block of an example web-service app */
mappings {
  path("/switches") {
    action: [GET: "listSwitches"] }
  path("/switches/:command") {
    action: [PUT: "updateSwitches"] }
}
def listSwitches() {
    switches.each {
      resp << [name: it.displayName, value: 
              it.currentValue("switch")]} //tainted
    return resp
}
def updateSwitches() {...}
/* An code block of an app using closures */
def someEventHandler(evt) {
    def currSwitches = switches.currentSwitch  //tainted
    def onSwitches = currSwitches.findAll {    //tainted
        switchVal -> switchVal == "on" ? true : false
    } 
}
/*  Implicit flows in an example app */
def batteryHandler(evt) {
  def batLevel = event.device?.currentBattery;
    if (batLevel < 25) {
	  switches.off()
	  def message = "battery low for device"
	  sendSMS(phone, message) 
    }
}
\end{lstlisting}

\begin{figure*}[!ht]
	\centering{\includegraphics[width=\textwidth]{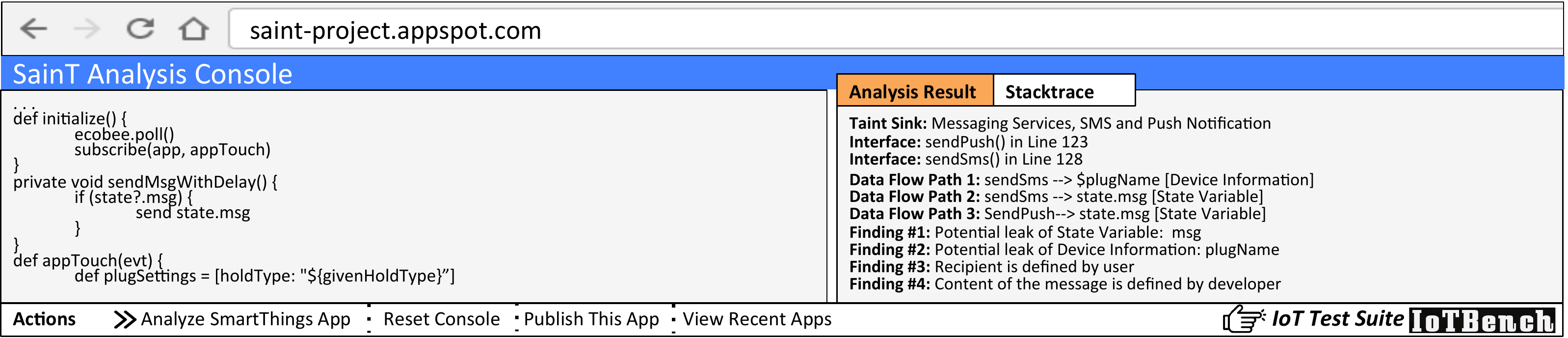}}
	\caption{Our \SainT data flow analysis tool designed for IoT apps. The left region is the analysis frame, and the right region is the output of an example IoT app for a specific data flow evaluation.\vspace{-2.5pt}}
	\label{fig:webapp}
\end{figure*}

\noindent\textbf{Call by Reflection.} The Groovy language supports programming by reflection (using the \texttt{GString} feature) \cite{SmartThingsAPI}, which allows a method to be invoked by providing its name as a string. For example, a method \texttt{foo()} can be invoked by declaring a string \texttt{name="foo"} and thereafter called by reflection through \texttt{\$name}; see Listing~\ref{listing-general} (lines 10--19) for another example. This can be exploited if an attacker can control the string used in call by reflection~\cite{fernandes2016security}, \eg if the code has \texttt{name=httpGet(URL)} and the URL is read from an external server. While SmartThings does not recommend using reflective calls, our study found that ten apps in our corpus use this feature (see Sec.~\ref{sec:evaluation}). To handle calls by reflection, \SainT's call graph construction adds all methods in an app as possible call targets, as a safe over-approximation. For the example in Listing~\ref{listing-general}, \SainT adds both \texttt{foo()} and \texttt{bar()} methods to the targets of the call by reflection in the call graph.

\noindent\textbf{Web Service Applications.} A web-service SmartThings app allows external entities to access smart devices and manage those devices. Such apps declare mappings relating \emph{endpoints}, HTTP operations, and callback methods. Listing~\ref{listing-general} (lines 20--33) presents a code snippet of a real web service app. The \texttt{/switches} endpoint handles an HTTP GET request that returns the state information of configured switches by calling the \texttt{listSwitches()} method; the \texttt{/switches/:command} endpoint handles a PUT request that invokes the \texttt{updateSwitches()} method to turn on or off the switches. The first prototype of \SainT did not flag the web service apps for leaking sensitive data. However, our manual investigation showed that the web-service apps respond to HTTP GET, PUT, POST, and DELETE requests from external services and may leak sensitive data. To correct this, we modified the taint-tracking algorithm to analyze what call back methods are declared through  the \texttt{mappings} declaration keyword~\cite{SmartThingsWebService}. Sensitive data leaked through those call back methods are then flagged by \SainT.
  
\noindent\textbf{Closures and Groovy-Specific Operations.} The Krushke sandbox enforced in SmartThings allows for closures and other Groovy-specific operations such as array insertions via $<<$. The SmartThings official developer guideline~\cite{smartThings-documentation} imposes certain restrictions on these operations. For instance, closures are disallowed outside of methods. \SainT's implementation follows the guideline and imposes the same restrictions. For closures, we found that apps often loop through a list of devices and use a closure to perform computation on each device in the list. Listing~\ref{listing-general} (lines 34--40) shows an example in which a closure is used to iterate through the \texttt{currSwitches} object to identify those switches that are turned on. For correct taint tracking, \SainT analyzes the structure of closures and inspects expressions in the closures to see how taints should be propagated.

\subsubsection{Implicit Flows}
\label{sec:implicit-flows}
An implicit flow occurs if the invocation of a sink interface is control dependent on a sensitive test used in a conditional branch. \SainT implements an algorithm designed to track implicit flows~\cite{kang2011dta}. It checks the condition of a conditional branch and sees whether it depends on a tainted value. If so, it taints all elements in the conditional branch~\cite{myers1999jflow}. Listing~\ref{listing-general} (lines 41--49) presents an example app, in which an implicit flow happens because a \texttt{sendSMS()} call is control dependent on a test that involves sensitive data \texttt{batLevel}. We found that IoT apps often use tainted values in control flow dependencies. In our analysis, approximately two-thirds of analyzed apps implement device actions (such as unlocking a door) in branches whose tests are based on tainted values (such as a user's presence). We leave the detection of implicit flows optional in \SainT, and evaluate the impact of implicit flow tracking on false positives in Sec.~\ref{sec:implicitFlowsEval}.

\subsection{Implementation}
\label{sec:saint_implementation}
The IR construction from the source code of the input IoT app requires the building of the app's ICFG. \SainT's IR-building algorithm directly works on the Abstract Syntax Tree (AST) representation of Groovy code. The Groovy compiler supports customizing the compilation process by supporting compiler hooks, through which one can insert extra passes into the compiler (similar to the modular design of the LLVM compiler~\cite{llvm}). The \SainT analyzer visits AST nodes at the compiler's semantic analysis phase where the Groovy compiler performs consistency and validity checks on the AST. Our implementation uses an \texttt{ASTTransformation} to hook into the compiler,  \texttt{GroovyClassVisitor} to extract the entry points and the structure of the analyzed app, and \texttt{GroovyCodeVisitor} to extract method calls and expressions inside AST nodes~\cite{GroovyVisitor}. This allows our implementation to use AST visitors to analyze expressions and statements, and get all necessary information to build IR.  

\SainT's taint analysis also uses Groovy AST visitors. It extends the \texttt{ASTBrowser} class implemented in the Groovy Swing console, which allows a user to enter and run Groovy scripts~\cite{groovy-console}. The implementation hooks into the IR of an app in the console and dumps information to the \texttt{TreeNodeMaker} class; the information includes an AST node's children, parent, and all properties built at the pre-defined compilation phase. This allows us to acquire the full AST including the resolved classes, static imports, the scope of variables, method calls,  interfaces accessed in an app. \SainT then uses Groovy visitors to traverse the IR's ICFG and perform taint tracking on it.

Since Groovy is a JVM-hosted language, one natural approach would be first to compile Groovy code into Java bytecode using the Groovy compiler and then build the IR via the help of the Soot analysis framework~\cite{vallee1999soot}. However, this approach was not feasible due to the heavy use of reflection in the bytecode generated by the Groovy compiler. In particular, the Groovy compiler translates every direct method call into a call by reflection. For instance, the example app in Fig.~\ref{fig:motExample} is compiled to bytecode with twelve reflective calls. Soot, unfortunately, does not produce good analysis results when the input bytecode uses reflection, as our experience suggests.

\noindent\textbf{Output of \SainT.} Fig.~\ref{fig:webapp} presents the screenshot of \SainT's analysis result on a sample app. A warning report by \SainT contains the following information: (1) full data flow paths between taint sources and sinks, and (2) the taint labels of sensitive data, and (3) taint sink information, including the hostname or URL, and contact information. 

\begin{table*}[t!]
\small
\def\arraystretch{1.05}
\setlength{\tabcolsep}{4pt}
\centering
\resizebox{1\textwidth}{!}{%
\begin{threeparttable}[b]
\begin{tabular}{lccccccccc}
  & \textbf{Official}\tnote{\textdagger} & \textbf{Third party} & \multicolumn{5}{c}{\textbf{Taint Sources}} &  \multicolumn{2}{c}{\textbf{Taint Sinks}}\\ \hline
\multicolumn{1}{|l|}{\textbf{App functionality}} & \multicolumn{1}{c|}{\textbf{Nr.}} & \multicolumn{1}{c|}{\textbf{Nr.}} &\textbf{Device State} & \textbf{Device Info\tnote{\textdagger}} &\textbf{Location} & \textbf{User Inputs} & \multicolumn{1}{c|}{\textbf{State Var.}} & \textbf{Internet} & \multicolumn{1}{c|}{\textbf{Messaging}}  \\ \hline
Convenience         & 80   & 26  & 96.2\%  & 87.7\%   & 51.9\%   & 97.2\% & 43.4\% & 25.5\% & 43.4\% \\
Security and Safety & 19   & 10  & 100\%   & 100\%    & 37.9\%   & 100\%  & 31.0\% & 3.4\%  & 86.2\% \\
Personal Care       & 10   & 0   & 90.0\%  & 60.0\%   & 50.0\%   & 90.0\% & 60.0\% & 20.0\% & 70.0\% \\
Home Automation     & 48   & 24  & 98.6\%  & 77.8\%   & 55.6\%   & 100\%  & 52.8\% & 8.3\%  & 40.3\% \\
Entertainment       & 10   & 0   & 90.0\%  & 70.0\%   & 70.0\%   & 100\%  & 60.0\% & 20.0\% & 10.0\% \\
Smart Transport     & 1    & 2   & 100\%   & 100\%    & 66.7\%   & 100\%  & 66.7\%& 33.3\% & 66.7\% \\ \hline
\textbf{Total}      & 168  & 62  &         &          &          &        &        \\
\end{tabular}

\begin{tablenotes}
    \item[\textdagger] Ten official apps and one third-party app do not request permission to devices, yet SmartThings explicitly grants access to device information such as hub id and manufacturer name (not shown).
\end{tablenotes}
\end{threeparttable}
}
\caption{Applications grouped by permissions to taint sources and sinks. App functionality shows the diversity of studied apps.}
\label{table:analysis}
\end{table*}

\section{Application Study}
\label{sec:evaluation}
This section reports our experience of applying \SainT on SmartThings apps to analyze how 230 IoT apps use privacy-sensitive data. Our study shows that approximately two-thirds of apps access a variety of sensitive sources, and 138 of them send sensitive data to taint sinks including the Internet and messaging channels. We also introduce an IoT-specific test suite called \iotbench. The test suite includes 19 hand-crafted malicious apps that are designed to evaluate taint analysis tools such as \SainT. We next present our taint analysis results by focusing on several research questions: 

\begin{enumerate}[leftmargin=2.5em, nolistsep]
    \item[\textbf{RQ1}] What are the potential taint sources whose data can be leaked?, and what are the potential taint sinks that can leak data? (Sec~\ref{sec:data-flow-analysis})
    \item[\textbf{RQ2}] What is the impact of implicit flow tracking on false positives? (Sec.~\ref{sec:implicitFlowsEval})
    \item[\textbf{RQ3}] What is the performance of \SainT in terms of precision and recall on \iotbench apps? (Sec.~\ref{sec:iotbench})
\end{enumerate}

\noindent\textbf{Experimental Setup.} In late 2017, we obtained 168 \emph{official} apps from the SmartThings GitHub repository~\cite{Official} and 62 community-contributed \emph{third-party} apps from the official SmartThings community forum~\cite{Community}. Table~\ref{table:analysis} categorizes the apps along with their requested permissions at install time. We determined the functionality of an app by checking its category in the SmartThings online store and also the definition block in the app's source code implemented by its developer. For instance, the ``entertainment'' category includes an app to control a device's speaker volume. We studied each app by downloading the source code and running analysis with \SainT. The official and third-party apps grant access to 49 and 37 ``different'' device types, respectively. The analyzed apps often implement SmartThings and Groovy-specific properties. Out of 168 official apps, \SainT flags nine apps using call by reflection, 74 declaring state variables, 37 implementing closures, and 23 using the OAuth2 protocol; out of 62 third-party apps, the results are one, 34, nine, and six, respectively. \SainT identifies when sensitive information transmits out via the internet and messaging services.

\noindent\textbf{Performance.} We assess the performance of \SainT on 230 apps. It took less than 16 minutes to analyze all apps. The experiment was performed on a laptop computer with a 2.6GHz 2-core Intel i5 processor and 8GB RAM, using Oracle’s Java Runtime 1.8 (64 bit) in its default settings. The average run-time for an app was 23$\pm$5 seconds.

\subsection{Data Flow Analysis}
\label{sec:data-flow-analysis}
In this subsection, we report experimental results of tracking explicit ``sensitive'' data flows by \SainT in IoT apps (implicit flows are considered in Section~\ref{sec:implicitFlowsEval}). Table~\ref{table:sendsSensitiveData} summarizes data flows via Internet and messaging services reported by \SainT. It flagged 92 out of 168 official and 46 out of 62 third-party apps have data flows from taint sources to taint sinks. We manually checked the data flows and verified that all reported ones are true positives. The manual checking process was straightforward to perform since the SmartThings apps are comparatively smaller than the apps found in other domains such as mobile phone apps.  Finally, although user inputs and state variables may over-approximate sources of sensitive information, during manual checking we made sure the reported data flows do include sensitive data.  
%The official apps have 244 lines of code on average, and the largest app has 2633 lines of code. Third-party apps have 247 lines of code on average, and the largest app has 1360 lines of code.

\SainT labels each piece of flow information with the sink interface, the remote hostname, and the URL if the sink is the Internet, and contact information if the sink is a messaging service. In Table~\ref{table:sendsSensitiveData}, the Internet column lists the number of apps that include only the taint source of the Internet. The Messaging column lists the number of apps that include only the taint source of some messaging service. 71.8\% of the analyzed apps are configured to send an SMS message or a push notification. As shown in the table, 47.2\% more apps include taint source in messaging services than the Internet. Finally, the Both column lists the number of apps (3.6\% of apps) that includes a taint source through both the Internet and messaging services.

\begin{table}[t!]
\centering
\def\arraystretch{1.2}
\setlength{\tabcolsep}{10pt}
\resizebox{\columnwidth}{!}{%
\begin{tabular}{@{}lcccc@{}}
\toprule
\textbf{Apps} & \bf{Nr}. & \bf{Internet} & \bf{Messaging} & \bf{Both}\\ \midrule \midrule
Official & 92 & 24 (26.1\%) & 63 (68.5\%) & 5 (5.4\%) \\ \hline
Third-party& 46 & 10 (21.7\%) & 36 (78.3\%) & 0 (0\%)  \\ \hline  \hline
\textbf{Total}  & 138 & 34 (24.6\%) & 99 (71.8\%) & 5 (3.6\%) \\ %\bottomrule
\end{tabular}}
\caption{Number of apps sending sensitive information through Internet and Messaging taint sinks.}
\label{table:sendsSensitiveData}
\end{table}

\begin{figure}[t]
    \centering{\includegraphics[width=1\columnwidth]{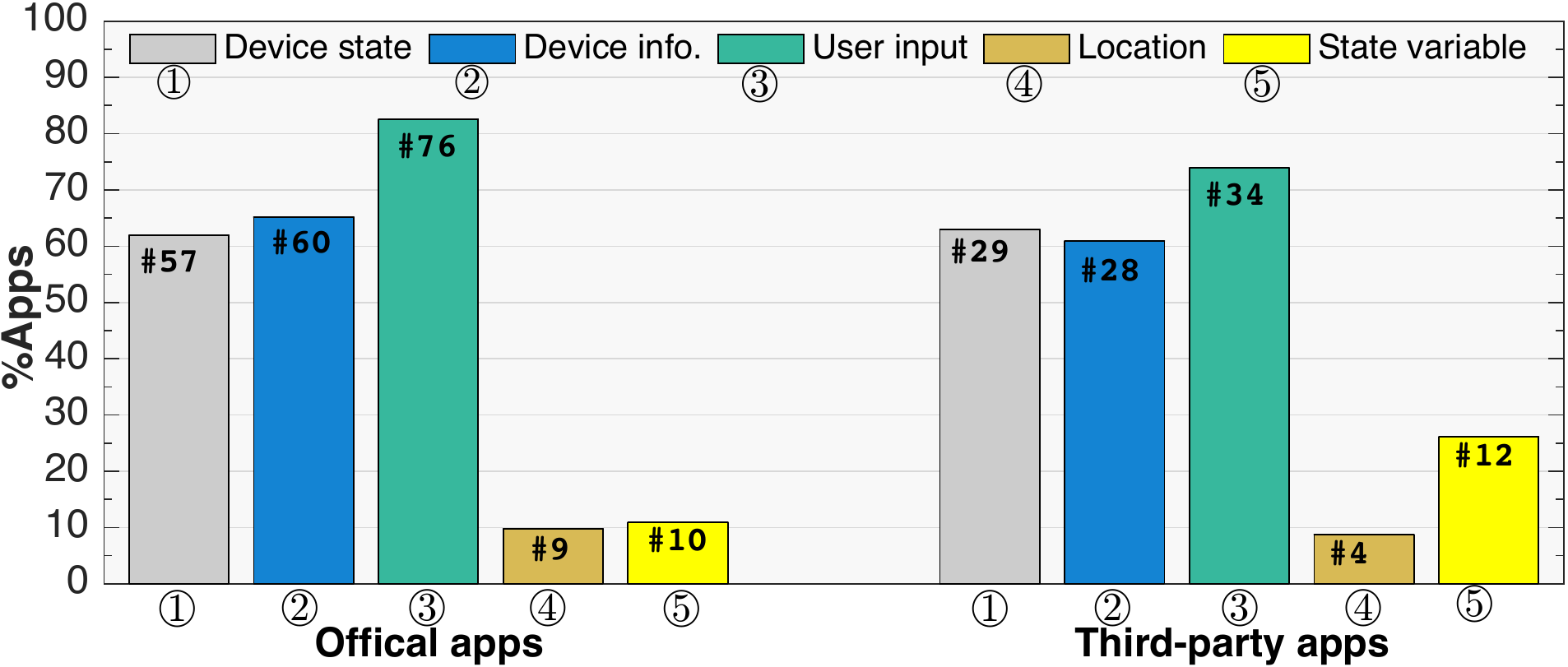}}
    \caption{Percentages of apps sending sensitive data for specific kinds of taint sources. The absolute numbers of apps are also presented after the \# symbol.}
    \label{fig:sensitive_source_leaks}
\end{figure}

\noindent\textbf{Taint Source Analysis.} Fig.~\ref{fig:sensitive_source_leaks} shows the percentages of apps that have sensitive data flows of a specific kind of taint sources. To measure this, we used sensitive data's taint labels provided by \SainT, which precisely describe what sources the data comes from. More than half of the apps send user inputs, device states, and device information. Approximately, one-ninth of the apps expose location information and values in state variables. We found that 64 out of 92 official apps and 30 out of 46 third-party apps send multiple kinds of data (\eg both device state and location information).

To better characterize the taint sources, we present the types of taint sources flagged by \SainT for apps that sends data in Table~\ref{fig:grid_sensitivity}. There are 92 official apps that send sensitive data, marked with ``O1 to O92", and 46 third-party apps that send sensitive data, marked with ``T1 to T46''. Out of 92 official apps, 28 apps (O1-O28) send one single kind of sensitive data, 16 apps (O29-O44) send two kinds of sensitive data, and the remaining 48 apps (O45-O92) send more than two and at most four kinds of sensitive data. Similar results also identified for third-party apps. Our investigation suggests that apps at the top of the Table~\ref{fig:grid_sensitivity} implement simpler tasks such as managing motion-activated light switches; the apps at the bottom tend to manage and control more devices to perform complex tasks such as automating many devices in a smart home. However, data flows depend on the functionality of the apps. For instance, a security and safety app managing few devices may send more types of sensitive data than an app designed for convenience that manages many devices.

\begin{figure}[t]
    \centering{\includegraphics[width=1\columnwidth]{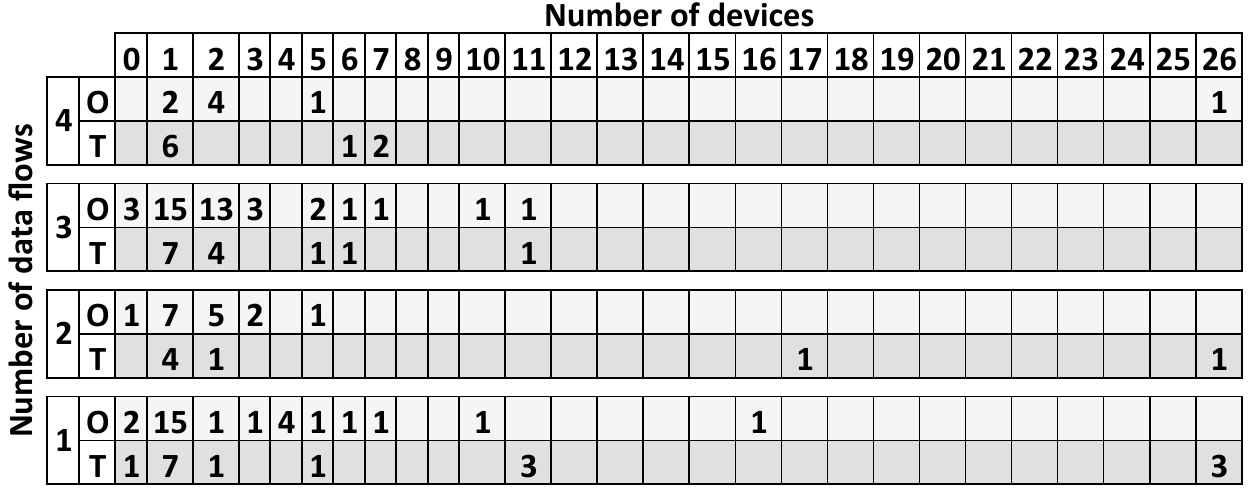}}
    \caption{The number of devices vs. the number of data flows based on taint labels in official (O) and third-party (T) apps. The numbers in the grids show the frequency of the apps.}
    \label{fig:devicesVSLeaks}
\end{figure}

In general, we found that there is no close relationship between the number of devices an app manages and the number of sensitive data flows. Fig.~\ref{fig:devicesVSLeaks} shows the number of apps for each combination of device numbers and numbers of data flows. As an example, there are two apps that manage seven devices and have four data flows. As shown in the figure, 15 official apps with a single device have three data flows, while an app with 16 devices has a single data flow. Similar results hold for third-party apps. Out of 46 third-party apps, 16 apps (T1-T16) have a single data flow, and the remaining 30 apps (T17-T46) have two to four data flows.

\begin{table}[th!]
\centering{\includegraphics[width=0.97\columnwidth]{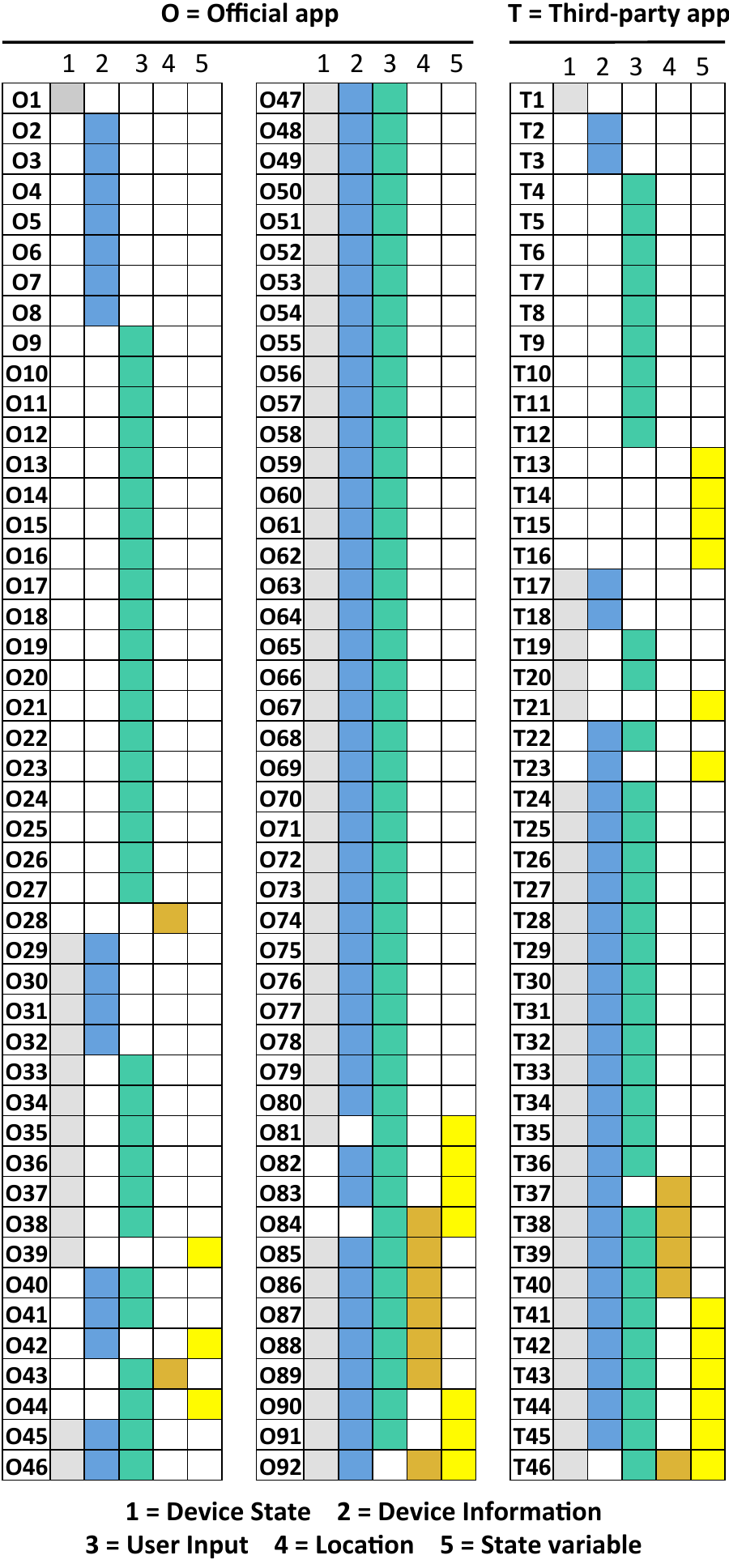}}
\caption{Data flow behaviour of each official (O1-O92) and third-party (T1-T46) app. 43.2\% of the official and 25.8\% of the third-party apps do not send sensitive data (not shown).}
\label{fig:grid_sensitivity}
\end{table}

\noindent\textbf{Taint Sink Analysis.} For a data flow, \SainT reports the interface name and the recipient (contact information, remote hostname or URL) defined in a taint sink. We use this information to analyze the number of different (a) sink interfaces and (b) recipients defined in each app. For (a), we consider apps invoking the same sink interface such as \texttt{sendSMS()} multiple times a single data flow, yet \texttt{sendNotification()} is considered a different interface from \texttt{sendSMS()}. We note for taint sink analysis we have a more refined notion of sinks than just distinguishing between the Internet and the messaging services; in particular, we take into account 11 Internet and seven messaging interfaces defined in SmartThings (see Appendix~\ref{appendix:sources-sinks}). For (b), we report the number of different recipients in invocations of sink interfaces used in an app.

\begin{figure}[t]
\centering
\begin{subfigure}[b]{0.24\textwidth}
  \centering
   \includegraphics[width=0.98\linewidth]{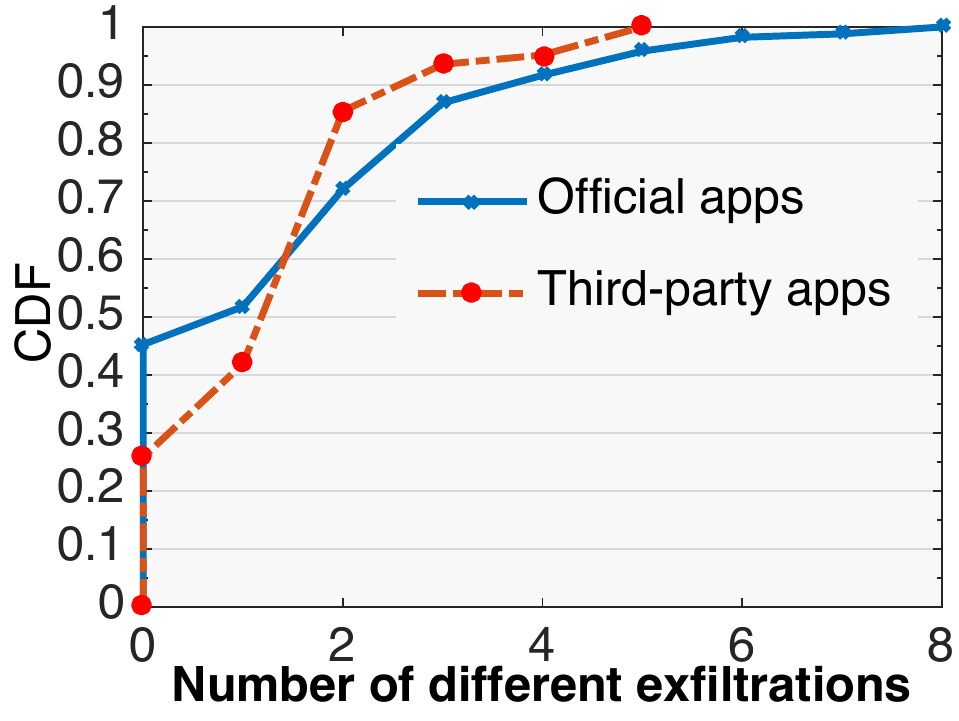}
  \caption{ }
  \label{fig:different-sinks}
\end{subfigure}%
\begin{subfigure}[b]{0.24\textwidth}
  \centering
  \includegraphics[width=0.98\linewidth]{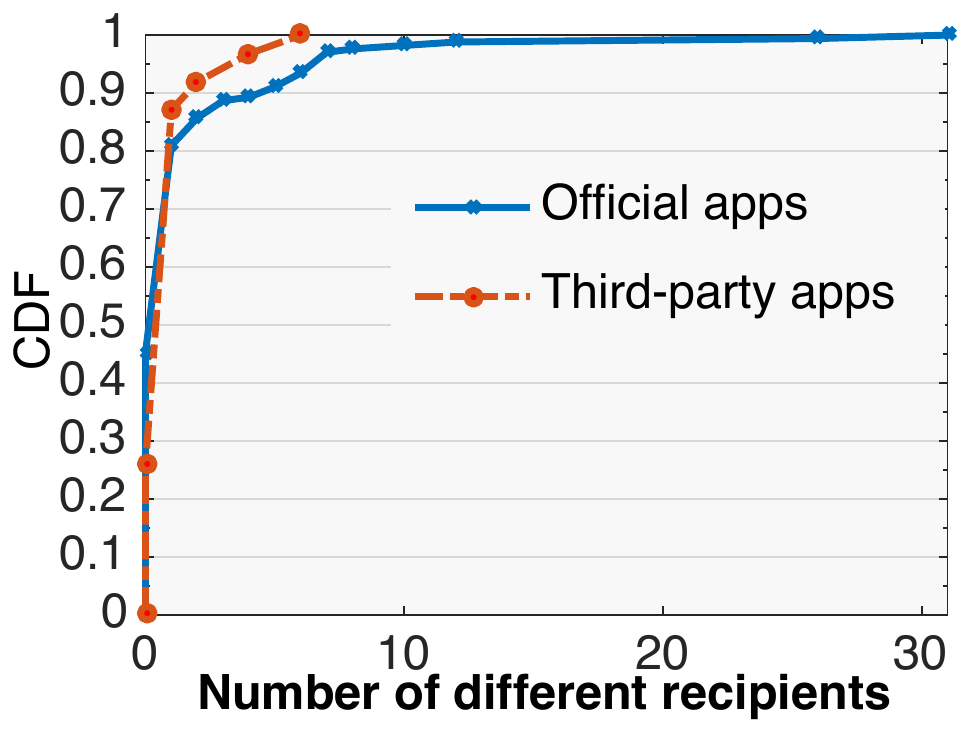}
  \caption{ }
  \label{fig:sink-recipients}
\end{subfigure}
\caption{Cumulative Distribution Function (CDF) of the number of different (a) sink interfaces and (b) recipients (contact information, remote hostname or URL) identified by \SainT.}
\label{fig:sink-analysis}
\end{figure}

A vast majority of apps contain data flows through either a push notification or an SMS message or makes a few external requests to integrate external devices with SmartThings. Fig.~\ref{fig:different-sinks} presents the CDF of the different sinks defined in official and third-party apps. Approximately, 90\% of the official apps contain at most four, and 90\% of the third-party apps contain at most three different invocations of sink interfaces (including apps that do not invoke sink interfaces). We also study the recipients at each taint sink reported in an app by \SainT. We first get the contact information for messaging, and hostname and URL for the Internet sinks. We then collect different contact addresses and URL paths to determine the recipients. Fig.~\ref{fig:sink-recipients} shows the CDF of the number of recipients defined in apps. The vast majority of apps involve a few recipients; they typically send SMS and push notifications to recipients. Approximately, 90\% of the official apps have less than three sink recipients, and  90\% of the third-party apps define at most two different recipients (including apps that do not implement taint sinks). A large number of recipients observed in official apps respond to external HTTP requests. For instance, a web-service app connects to a user's devices, accesses their events and commands, and uses their state information to perform actions, and an app allows users to stream their device events to a remote server for data analysis and visualization. This leads to using a variety of taint sinks and URLs to access and manage various devices.

\noindent\textbf{Recipient and Content Analysis.} When a data transmitted over a sink-interface call, \SainT reports who defines the recipient and the content in the data flow. With content and recipient analysis, we can have a refined understanding of who defines the recipient and content. In particular, this helps identify if the recipient is authorized by a user, if sensitive data is sent to a legitimate or malicious external server, and if the app conforms its functionality. The recipient refers to who receives the message in a messaging service or who is the destination of an Internet communication. The content refers to the message used in a messaging service or the parameter of a request (e.g., HTTP GET or PUT) used in an Internet communication. For instance, a call to the \texttt{sendSMS()} interface requires the phone number as the recipient and also a message to that recipient. We extended \SainT to output whether the recipient and the content of a sink-interface call are specified by a \emph{user} at install time, by a \emph{developer} via hard-coded strings in an app's source code, or by an \emph{external} entity such as a remote server (in this case, a remote server first sends the recipient information, and then the app sends sensitive data to the recipient).  

Table~\ref{table:whowhat_analysis} presents the number of times a user, a developer, or an external party specifies the recipient or the content used in a data flow. The messaging rows of the table tell that, in official apps, users specify recipients 154 times, while contents are specified by users five times and 149 times by developers; for third-party apps, users define recipients 67 times, while message contents are specified by users five times and 63 times by developers. In contrast, message contents are often hard-coded in the apps by developers. Table~\ref{table:whowhat_analysis} shows a different story for Internet-sink calls. In this case, recipients and contents are often specified by developers and external services. An app in which recipients and contents of Internet-sink call are specified by external services is often a web-service app. As detailed in Sec.~\ref{sec:advanced-taint-tracking}, web-service apps expose endpoints and respond to requests from external services. These apps allow external services to access and manage devices. Additionally, in some apps, developers hard-code the recipients and contents of Internet communications to send information to external remote servers. 

\begin{table}[t!]
\centering
\def\arraystretch{1}
\setlength{\tabcolsep}{0.9pt}
\label{my-label}
\resizebox{\columnwidth}{!}{
\begin{tabular}{ll|c|c|c|c|c|c|}
\cline{3-8}
 &  & \multicolumn{6}{c|}{\textbf{Taint sink analysis}} \\ 
\cline{3-8}
 &  & \multicolumn{3}{c|}{\textbf{Recipient defined by}} & \multicolumn{3}{c|}{\textbf{Content defined by}} \\ \hline 
\multicolumn{1}{|l|}{\textbf{Taint Sinks}} & \textbf{Apps} & \textbf{User} & \textbf{Developer} & \textbf{External} & \textbf{User }& \textbf{Developer} & \textbf{External} \\ \hline \hline
\multicolumn{1}{|l|}{\multirow{2}{*}{Messaging}} & Official & 154 & 0 & 0 & 5 & 149 & 0 \\ \cline{2-8} 
\multicolumn{1}{|l|}{} & Third-party & 67 & 0 & 0 & 4 & 63 & 0 \\ \hline \hline
\multicolumn{1}{|l|}{\multirow{2}{*}{Internet}} & Official & 2 & 48 & 44 & 0 & 54 & 40 \\ \cline{2-8} 
\multicolumn{1}{|l|}{} & Third-party & 0 & 13 & 12 & 0 & 13 & 12 \\ \hline
\end{tabular}}
\caption{Recipient and content analysis of data flows.}
\label{table:whowhat_analysis}
\end{table}

\noindent\textbf{Summary.} Our study of 168 official and 62 third-party SmartThings IoT apps shows the effectiveness of \SainT in accurately detecting sensitive data flows. \SainT flagged 92 out of 168 official apps, and 46 out of 62 third-party apps transmit at least one kind of sensitive data over a sink-interface call. We analyzed reported data's taint labels provided by \SainT, which precisely describe the data source. Using this information, we found that half of the analyzed apps transmit at least three kinds of sensitive data. We used sink interface names and recipients to analyze the number of different Internet and messaging interfaces and recipients in an app. Approximately, two-thirds of the apps define at most two separate sink interfaces and recipients. Moreover, we extended our analysis to identify whether the recipient and the content of a sink-interface call are specified by a user, a developer, or an external entity. All recipients of messaging-service calls are defined by users and approximately nine-tenths of message contents are defined by developers. For Internet sinks, nine-tenths of the Internet recipients and contents are specified by developers or external servers. 

\SainT's findings provide a means to automatically detect and evaluate sensitive data flows. Where intentional, developers and device manufacturers can provide explanations and warnings about discovered sensitive flows through system documentation or other means. Where unintentional or malicious, device implementations can be rejected or modifications required.

\subsection{Implicit Flows} 
\label{sec:implicitFlowsEval}
We repeated our experiments by turning on both explicit and implicit flows tracking. Approximately two-thirds of the apps invoke some sink interface that is control dependent on sensitive tests. However and somewhat surprisingly, there are only six extra warnings produced when turning on implicit flows. The reason we found is that most of those sink calls already leak data through explicit flows. For example, in one app, \texttt{x} gets the state of a device \texttt{x=currentState("device")} and, when a user is present, \texttt{x} is sent out via an SMS message; even though there is an implicit flow (because sending the message depends on whether the user is present), there is also an explicit flow as the device information is sent out. The six extra warnings are all about sending out hard-coded strings: ``Your mail has arrived!'', ``Your ride is here!'', ``No one has fed the dog'', ``Remember to take your medicine'', ``Potential intruder detected'', and ``Gun case has moved!''. These messages contain information in themselves and are sent conditionally upon sensitive information; therefore we believe information is indeed leaked in these cases. We note that turning on implicit flow tracking increases the tracking overhead as more identifiers need to be tracked; however, based on the results, turning on implicit flow tracking on SmartThings IoT apps does not lead to an unmanageable number of false positives.

\subsection{IoTBench}
\label{sec:iotbench}
We introduce an IoT-specific test suite, an open repository for evaluating information leakage in IoT apps. We designed our test suite similar to those designed for mobile systems~\cite{ArztFlowdroid, EnckTaintDroid} and the smart grid~\cite{mclaughlin2012sabot}; they have been widely adopted by the security community. \iotbench currently includes 19 hand-crafted malicious SmartThings apps that contain data leaks. Sixteen apps have a single data leak, and three have multiple data leaks; a total of 27 data leaks via either Internet and messaging service sinks. We carefully crafted the \iotbench apps based on official and third-party apps. They include data leaks whose accurate identification through program analysis would require solving problems including multiple entry points, state variables, call by reflection, and field sensitivity. Each app in \iotbench also comes with ground truth of what data leaks are in the app; this is provided as comment blocks in the app's source code. \iotbench can be used to evaluate both static and dynamic taint analysis tools designed for SmartThings apps; It enables assessing a tool's accuracy and effectiveness through the ground truths included in the suite. We present three example SmartThings apps and their privacy violations in Appendix~\ref{appendix:iotbench_apps}.  We made \iotbench publicly available as well:
\begin{center}
\url{https://github.com/IoTBench/}. 
\end{center}

\noindent\textbf{\SainT results on \iotbench.} We next report the results of using \SainT on 19 \iotbench apps. In the discussion, we will use app IDs defined in Table~\ref{appendix:iotbench-table} in Appendix~\ref{appendix:iotbench_apps}. \SainT produces false warnings for two apps that use call by reflection (apps 6 and 7). These two apps invoke a method via a string. \SainT over-approximates the call graph by allowing the method invocation to target all methods in the app. Since one of the methods leaks the state of a door (locked or unlocked) to a malicious URL and the mode of a user (away or home) to a hard-coded phone number, \SainT produces warnings. However, it turns out that the data-leaking method would not be called by the reflective calls in those two apps. This pattern did not appear in the 230 real IoT apps we discussed earlier. \SainT did not report leaks for two apps that leak data via side channels (apps 18 and 19). For example, in one app, a device operates in a specific pattern to leak information. As our threat model states, data leaks via side channels are out of the scope of \SainT and are not detected.

\section{Limitations and Discussion}
\label{sec:limit-discuss}
\SainT leaves detecting implicit flows optional. Even though our evaluation results on SmartThings apps show that tracking implicit flows does not lead to over-tainting and false positives, whether this holds on apps of other IoT platforms and domains would need further investigation. Another limitation is \SainT's treatment of call by reflection. As discussed in Sec.~\ref{sec:methodology}, it constructs an imprecise call graph that allows a call by reflection target any method. This increases the number of methods to be analyzed and may lead to over-tainting. We plan to explore string analysis to statically identify possible values of strings and refine the target sets of calls by reflection.

\SainT treats all user inputs and state variables as taint sources, even though some of those may not contain sensitive information. However, this has not led to false positives in our experiments. Another limitation is about sensitive strings. An app may hard code a string such as ``Remember to take your Viagra in the cabinet'' and send the string out.  Though the string contains sensitive information, \SainT does not report a warning (unless there is an implicit flow and implicit flow tracking is turned on). Determining whether hard-coded strings contain sensitive information may need user help or language processing.

Finally, \SainT's implementation and evaluation are purely based on the SmartThings programming platform designed for home automation. There are other IoT domains suitable for studying sensitive data flows, such as FarmBeats for agriculture~\cite{vasisht2017farmbeats}, HealthSaaS for healthcare~\cite{healthIoT}, and KaaIoT for the automobile~\cite{Kaa}. We plan to extend \SainT's algorithms designed for SmartThings to these platforms and identify sensitive data flows.

\section{Related Work}
\label{sec:related}
There has been an increasing amount of recent research exploring IoT security. These works centered on the security of emerging IoT programming platforms and IoT devices. For example, Fernandes et al.~\cite{fernandes2016security} identified design flaws in permission controls of SmartThings home apps and revealed the severe consequences of over-privileged devices. In another paper, Xu et al.~\cite{xu2014security} surveyed the security problems on IoT hardware design. Other efforts have explored vulnerability analysis within specific IoT devices~\cite{oluwafemi2013experimental, ho2016smart}. These works have found that apps can be easily exploited to gain unauthorized access to control devices and leak sensitive information of users and devices. 

Many of previous efforts on taint analysis focus on the mobile-phone platform~\cite{EnckTaintDroid, ZhuTaint, GuTaint, clause2007dytan, ArztFlowdroid, gordon2015information}. These techniques are designed to model domain-specific challenges like on-demand algorithms for context and object sensitivity. Several efforts on IoT analysis have focused on the security and correctness of IoT programs using a range of analyses. To restrict the usage of sensitive data, FlowFence~\cite{fernandes2016flowfence, rahmati2016applying} enforces sensitive data flow control via opacified computation. ContexIot~\cite{jia2017contexiot} is a permission-based system that infers the IoT app context automatically and to enforce permissions based on that context. In contrast, to our best knowledge, \SainT is the first system that that precisely detects sensitive data flows in IoT apps by carefully identifying a complete set of taint sources and taint sinks, adequately modeling IoT-specific challenges like app lifecycles, and event handler methods, and addressing platform- and language-specific problems.

\section{Conclusions}
\label{sec:concl}
One of the central challenges of existing IoT is the lack of visibility into the use of data by applications. In this paper, we presented \SainT \footnote{\SainT is openly available at \url{http://saint-project.appspot.com/}.}, a novel static taint analysis tool that identifies sensitive data flows in IoT apps. \SainT translates IoT app source code into an intermediate representation that models the app’s lifecycle--including program entry points, user inputs, events, and actions.  Thereafter we perform efficient static analysis tracking information flow from sensitive sources to sink outputs. We evaluated \SainT in two studies; a horizontal SmartThings market study validating \SainT and assessing current market practices, and a second study on our novel \iotbench app corpus. These studies demonstrated that our approach can efficiently identify taint sources and sinks and that most market apps currently contain sensitive data flows.

\SainT represents a potentially important step forward in IoT analysis, but further work is required.  In future work, we will expand our analysis to support more platforms as well as refine our analysis for more complex and subtle properties.  At a higher level, we will extend the kinds of analysis provided by the online systems and therein provide a suite of tools for developers and researchers to evaluate implementations and study the complex interactions between users and the IoT devices that they use to enhance their lives.  
Lastly, we will expand the \iotbench app suite.  In particular, we are studying the space of privacy violations reported in academic papers, community forums, and from security reports, and will reproduce unique flow vectors in sample applications.

\section{Acknowledgment}
Research was sponsored by the Army Research Laboratory and was accomplished under Cooperative Agreement Number W911NF-13-2-0045 (ARL Cyber Security CRA). The views and conclusions contained in this document are those of the authors and should not be interpreted as representing the official policies, either expressed or implied, of the Army Research Laboratory or the U.S. Government. The U.S. Government is authorized to reproduce and distribute reprints for Government purposes notwithstanding any copyright notation here on. This work is also partially supported by the US National Science Foundation (Awards: NSF-CAREER-CNS-1453647, NSF-1663051) and Florida Center for Cybersecurity (FC2)'s Capacity Building Program (Award\#: AWD000000007773).

\bibliographystyle{IEEEtran}
%\bibliography{references.bib}

\appendix{}

%reset counters
\setcounter{figure}{0}
\setcounter{table}{0}
\setcounter{lstlisting}{0}
\setcounter{section}{0}

\section{Source Code of the Example App}
\label{appendix:example-app}
We present the Groovy source code of the home-automation app's IR shown in Figure \ref{fig:motExample}, Section~\ref{sec:methodology}.

\begin{lstlisting}[caption=an example home-automation app,label=listing-sourceCode]
definition(
    name: "SmartApp",
    namespace: "mygithubusername",
    author: "SainT",
    description: "This is my SmartApp for my home automation",
    category: "My Apps",
    iconUrl: "https://s3.amazonaws.com/smartapp-icons/Convenience/Cat-Convenience.png",
    iconX2Url: "https://s3.amazonaws.com/smartapp-icons/Convenience/Cat-Convenience@2x.png",
    iconX3Url: "https://s3.amazonaws.com/smartapp-icons/Convenience/Cat-Convenience@2x.png")

preferences {
    section("When you are away/home") {
        input "presenceSensor", "capability.presenceSensor", multiple: true, 
        required: true, title: "Which presence sensor?"
    }

    section("Turn on the lights") {
        input "theSwitches", "capability.switch", required: true, multiple: true,
        title: "Which lights?"
    }

    section("Lock/Unlock door") {
        input "theDoor", "capability.door", multiple: false, 
        required: true, title: "Which door?"
    }

    section("Notify between what times?") {
        input "fromTime", "time", title: "From", required: true
        input "toTime", "time", title: "To", required: true
    }

    section("Send Notifications?") {
        input("recipients", "contact", title: "Send notifications to") {
            input "phone", "phone", title: "Warn security with text message",
            description: "Phone Number", required: true
        }
    }
}

def installed() {
    initialize()
}

def updated() {
    log.debug "Updated with settings: ${settings}"
    unsubscribe()
    initialize()
}

def initialize() {
    log.debug "initialize configured"
    subscribe(presenceSensor, "present", h1) 
    subscribe(presenceSensor, "not present", h2) 
}

def h1(evt) {
    log.debug "presence active called: $evt"
    x()
}

def h2(){
    log.debug "presence not active called: $evt"
    theSwitches.off()
    theDoor.unlock()
	
    def between  =  y()
    if (between){
        z()
    }

    def currSwitches = theSwitches.currentSwitch
    def onSwitches = currSwitches.findAll { switchVal ->
        switchVal == "on" ? true : false
    }
    log.debug "${onSwitches.size()} out of ${switches.size()} switches are on"
}

def x(){
    theSwitches.on()
    theDoor.unlock()
    def currSwitches = theSwitches.currentSwitch
    def onSwitches = currSwitches.findAll { switchVal ->
        switchVal == "on" ? true : false
    }
    log.debug "${onSwitches.size()} out of ${theSwitches.size()} switches are on"
}

def y(){
    log.debug "In time method"
	return timeOfDayIsBetween(fromTime, toTime, new Date(), location.timeZone)
}

def z(){
    log.debug "recipients configured: $recipients"
    sendSms(phone, "The ${theDoor.displayName} is locked and the ${theSwitches.displayName} is off!")
    def latestValue = theDoor.latestValue("door")
    log.debug "message sent, the door status is $latestValue"
}
\end{lstlisting}
\vspace{2em}

\section{IoTBench Apps}
\label{appendix:iotbench_apps}
Table~\ref{appendix:iotbench-table} presents \iotbench apps categorized by their data leak ground-truth. We present three example apps and their privacy violations below.

%App 11
Our first app ``Implicit Permission 1" (ID: 11) sends a short message to household members when everyone is away. We update an existing legitimate app to include a code block that sends the state of the door via the \texttt{leak()} method to a remote server (see Listing~\ref{listingMalware1}). A privacy violation occurs because it informs the server of the absence of household members and the door state.  

%belowskip=0\baselineskip
\begin{lstlisting}[caption=Device state leak via the internet interface, label=listingMalware1, belowskip=0.1\baselineskip]
if (everyoneIsAway()){
    //app logic
    leak() // invoke when everyone is away
}
def leak() {
    Params = [
    uri: "https://malicious-url", 
    body: ["condition":"$thedoor.latestValue("door")"]]
    httpPost(Params) // leak
}
\end{lstlisting}

%app 12
The second app ``Explicit-Implicit" (ID: 14) sends a short message to users when a door lock has a low battery.  A code block is added to an existing app to send the battery level (implicit permission) and hub id (explicit permission) to a third-party's phone number via \texttt{sendSms()} when the \texttt{sms\_send} variable is true (see Listing~\ref{listingMalware2}). Here, \texttt{sms\_send} is tainted via the \texttt{state} object's \texttt{SMS} field. The leaked battery level is a privacy violation. 

%belowskip=0\baselineskip
\begin{lstlisting}[caption=Data leak of battery level and hub id,label=listingMalware2, belowskip=0.1\baselineskip]
def BatteryPowerHandler(evt) {
    sms_send = state.SMS // set true 
    msg = "$doorBattery.currentValue("battery")
            power is out in hub  ${evt.hubId}!"
    sendPush(msg) // user gets a push notification 
    
    if (sms_send) { // attacker gets the same message
        sendSms(attacker_phone#, msg) // leak
     }
}
\end{lstlisting}

%app 5
Our final example is the ``Call by Reflection 1" app (ID: 5). The app is used to trigger the alarm when smoke is detected. This app obtains the method name string from a remote server and uses this string to invoke \texttt{\$state.method} (see Listing~\ref{listingMalware3}). Thus, the \texttt{updateApp()} method can be called by reflection. Because \SainT adds all methods in an app as possible call targets, it detects a data leak in \texttt{updateApp()}, which disables alarm by unsubscribing the ``smoke-detected" event and sends this information to a hardcoded phone number.

%belowskip=0\baselineskip
\begin{lstlisting}[caption=Data leak via a reflective call, label=listingMalware3, belowskip=0.1\baselineskip]
def attack(){
  httpGet("http://maliciousServer.com"){ 
    resp ->
        if(resp.status == 200){
            state.method = resp.data.toString()
        }
    "$state.method"() // reflective call
}           
updateApp() {
    unsubscribe() // revoke smoke detector events
    sendSMS("number","$detector is revoked")
}       
\end{lstlisting}

\section{Taint Source and Taint Sink APIs}
\label{appendix:sources-sinks}
We present SmartThings APIs that are taint sinks in Table~\ref{sinkAPI} and APIs that are taint sources in Table~\ref{sourceAPIs}. We refer the interested reader to SmartThings API documentation for the details~\cite{SmartThingsAPI}. For taint sinks, SmartThings recently announced asynchronous HTTP requests available as a beta development feature~\cite{smartThings-documentation}. However, the analyzed apps do not use asynchronous HTTP APIs; thus we exclude them from the list. 
We note that some taint-source APIs are used together with the device names assigned by the developer, or require specific device capabilities to use them. Therefore, the number of taint sources used in an app differs based on the app's context.

\makeatletter
    \setlength\@fptop{0\p@}
\makeatother

\begin{table}[ht!]
\centering
\def\arraystretch{1}
\setlength{\tabcolsep}{14pt}
{\small{
\begin{tabular}{@{}ll@{}}
\toprule
\multicolumn{1}{l}{\textbf{Internet}} & \multicolumn{1}{l}{\textbf{Messaging}} \\ \midrule\midrule
httpDelete()                          & sendSms()                              \\
httpGet()                             & sendSmsMessage()                       \\
httpHead()                            & sendNotificationEvent()                \\
httpPost()                            & sendNotification()                     \\
httpPostJson()                        & sendNotificationToContacts()           \\
httpPut()                             & sendPush()                             \\
httpPutJson()                         & sendPushMessage()                      \\
GET (web service apps)                &                                        \\
PUT (web services apps)               &                                        \\ 
POST (web service apps)               &                                        \\
DELETE (web service apps)             &                                        \\\bottomrule
\end{tabular}}}
\caption{SmartThings taint-sink APIs.}
\label{sinkAPI}
\end{table}

\begin{sidewaystable*}
\centering
\scriptsize
\def\arraystretch{1.1}
\begin{tabularx}{\textwidth}{p{4cm}p{7cm}p{4cm}p{7.5cm}}
\hline
\textbf{Name of the interface}             & \textbf{Definition}                                                       &      \textbf{Name of the interface}            & \textbf{Definition}                                     \\ \hline \hline
\multicolumn{2}{c}{\textbf{Device Information}}                                                                        &          \multicolumn{2}{c}{\textbf{Device State}}                                            \\ \hline
capabilitiy.$<$device type or attribute$>$ & Allow to abstract devices into their underlying capabilities              &    latestState()                              & Get the latest Device State record for the specified attribute    \\
getManufacturerName()                      & Get the manufacturer name of the device                                  &    statesSince()                              & Get a list of Device State since the date specified     \\
getModelName()                             & Get the model name of the device                                         &    getArguments()                             & Get the list of argument types for the command\\
getName()                                  & Get the internal name of the device, Hub, command, or attribute                   &    getDateValue()                             & Get the value of the event as a Date object   \\
GetupportedAttributes()                   & Get the list of device attributes                                        &    getDescriptionText()                       & Get the description of the event\\
GetupportedCommands()                     & Get the list of device commands                                          &    getDoubleValue()                           & Get the value of the event as a Double\\
hasAttribute()                             & Determine if the device has the specified attribute                      &    getFloatValue()                            & Get the value of the as a Float\\
hasCapability()                            & Determine if the device supports the specified capability                &    getIntegerValue()                          & Return the value of the event as an Integer\\
hasCommand()                               & Determine if the device has the specified command name                    &    getJsonValue()                             & Get the value of the event as a parsed JSON\\
latestValue()                              & Get the latest reported value for the specified attribute                &    getLastUpdated()                           & Get the last time the event was updated\\
getFirmwareVersionString()                 & Get the firmware version of the Hub device                               &    getLongValue()                             & Get the value of the event as a Long\\
getId()                                    & The unique system identifier for the device or the Hub                    &    getName()                                  & Get the name of the event\\
getLocalIP()                               & The local IP address of the Hub device                                    &    getNumberValue()                           & Get the value of the event as a number \\
getLocalSrvPortTCP()                       & The local server TCP port of the Hub device                               &    getNumericValue()                          & Get the value of the event as a number\\
getDataType()                              & Get the data type of the device attribute                                &    getUnit()                                  & Get the unit of measure for the event\\
getValues()                                & Get the possible values for the device attribute                         &    getValue()                                 & Get the value of the event as a String\\
getType()                                  & Get the type of the Hub device                                           &    getData()                                  & Get a map of any additional data on the event\\
getZigbeeId()                              & Get the ZigBee ID of the Hub                                             &    getDate()                                  & Acquisition time of the device state record\\
getZigbeeEui()                             & Get the ZigBee Extended Unique Identifier of the Hub                     &    getDescription()                           & The raw description that generated the event\\
events()                                   & Get a list of events for the Device in reverse chronological order        &    getDevice()                                & Get the device associated with the event\\
eventsBetween()                            & Get a list of events between the specified start and end dates            &    getDisplayName()                           & Get the user-friendly name of the source of the event\\
eventsSince()                              & Get a list of events since the specified date                             &    getDeviceId()                              & Unique identifer of the Device associated with the event\\
getCapabilities()                          & The list of capabilities provided by this Device                          &    getIsoDate()                               & Acquisition time of the event as an ISO-8601 String\\
getDeviceNetworkId()                       & Get the device network ID for the device                                 &    getSource()                                & The source of the event\\
getDisplayName()                           & The label of the device assigned by the user                              &    getXyzValue()                              & Value of the event as a 3-entry Map\\
getHub()                                   & The Hub associated with this device                                       &    isPhysical()                               & TRUE if the event is from a physical actuation of the device\\
getLabel()                                 & The name of the device in the mobile application or Web IDE               &    isStateChange()                            & TRUE if the attribute value for the event has changed\\
getLastActivity()                          & The date of the last event from the device                                &    isDigital()                                & TRUE if the event is from a digital actuation of the device\\
getManufacturerName()                      & Gets the manufacturer name of the device                                  &    currentState()                             & Gets the latest State for the specified attribute\\
getModelName()                             & Gets the model name of the device                                         &    currentValue()                             & Gets the latest reported values of the specified attribute\\
deviceName.capabilities                    & Gets the device capabilities                                              &    getStatus()                                & Gets the current status of the device                      \\
getTypeName()                              & The type of the device                                                    &                                               &                                                             \\ \hline
\multicolumn{2}{c}{\textbf{Location}}                                                                                  &    \multicolumn{2}{c}{\textbf{User Inputs}}                                                                 \\ \hline
getContactBookEnabled()                    & Determine if the Location has Contact Book enabled                       &    input ``someSwitch'', ``capability.switch''    & User preferences for the devices (accessed as \$someSwitch)\\
getCurrentMode()                           & Gets the current mode for the location                                    &    input ``someMessag'', ``text''                & User preferences for message (accessed as \$someMessage)\\
getId()                                    & Gets the unique internal system identifier for the location               &    input ``someTime'', ``time''                   & User preferences for the time (accessed as \$someTime)\\
getHubs()                                  & Gets the list of Hubs for the location                                    &    input ``someTime'', ``time''                  & User preferences for the time (accessed as \$someTime)\\
getLatitude()                              & gets the geographical latitude of the location                            &    input ``minutes'', ``time''                    & User preferences for time span (accessed as \$minutes)\\ \cline{3-4}
getLongitude()                             & Gets the geographical longitude of the location                           &    \multicolumn{2}{c}{\textbf{State Variables}}                                                    \\ \cline{3-4}
getMode()                                  & Gets the current mode name for the location                               &    state                                      & Defines the state variable state\\
setMode()                                  & Sets the mode for the location                                            &    atomicState                                & Defines the state variable atomicState\\
getTimeZone()                              & Gets the time zone for the location                                       &\\
getZipCode()                               & Gets the ZIP code for the location                                        &\\
getLocationId()                            & The unique identifier for the location associated with the event          &\\
getLocation()                              & The Location associated with the event                                    &\\ \hline
\end{tabularx}
\caption{SmartThings taint-source APIs. The complete list can be accessed in our project page [anon].}
\label{sourceAPIs}
\end{sidewaystable*}

\begin{table*}[t!]
\begin{minipage}{\textwidth}
\newcolumntype{P}[1]{>{\centering\arraybackslash}p{#1}}
\def\arraystretch{1.2}
\centering
\small
\begin{threeparttable}[b]
\begin{tabular}{|p{2.3cm}|p{3.2cm}p{7cm}P{1cm}|}
\hline
\textbf{App Category}  & \textbf{ID/App Name}  & \textbf{App Description\tnote{\textdaggerdbl}} & \textbf{Results\tnote{\textdagger}} \\ \hline\hline

\multirow{2}{*}{\textbf{Lifecycle}}    & 1- Multiple Entry Point 1   & The app stores different sensitive data under the same variable name in different functions and only one of them is leaked.  & \CheckmarkBold                  \\ \cline{2-4} 
                                         & 2- Multiple Entry Point 2   &  The app stores different sensitive data under the same variable name in different functions and more than one piece of data is leaked.  &  \CheckmarkBold         \\ \cline{2-4} 
                                         \hline
\multirow{1}{*}{\textbf{Field Sensitivity}} & 3- State Variable 1     &   A state variable in the \texttt{state} object's field  stores sensitive data. It is used in different functions and leaked  through various sinks.  &     \CheckmarkBold   \\ \cline{2-4}  \hline  
\multirow{1}{*}{\textbf{Closure}} & 4- Leaking via Closure &   A variable is tainted with the use of closures. The sensitive data is then leaked via different sinks.  &     \CheckmarkBold   \\ \cline{2-4}  \hline
\multirow{3}{*}{\textbf{Reflection}} & 5- Call by Reflection 1 & A string is  requested via \texttt{HttpGet} interface and used in a call by reflection. A method leaks device information. & \text{\sffamily O} \\ \cline{2-4} 
                                         &  6- Call by Reflection 2  & A string is used to invoke a method via call by reflection. A method leaks the state of a door.  &  \text{\sffamily X}  \\ \cline{2-4}
                                         &  7- Call by Reflection 3 & A string is used to invoke a method via call by reflection. A method leaks the mode of a user. & \text{\sffamily X} \\ \cline{2-4} \hline
\multirow{3}{*}{\textbf{Device Objects}} & 8- Multiple Devices 1 & Various sensitive data is obtained from different devices and leaked via different sinks.  & \CheckmarkBold \\ \cline{2-4}
                                         & 9- Multiple Devices 2 & Sensitive data from various devices is tainted and leaked via different sinks. & \CheckmarkBold \\ \cline{2-4}
                                         & 10- Multiple Devices 3 & A taint source is obtained from device states and information and leaked via messaging services. & \CheckmarkBold  \\ \cline{2-4} \hline
\multirow{4}{*}{\textbf{Permissions}} & 11- Implicit  1 & A malicious URL is                                               hard-coded and device states (implicit permission) are leaked via sinks using the hard-coded URL. & \CheckmarkBold \\ \cline{2-4}
                                        & 12- Implicit  2 & The contact information (\ie phone number) is hard-coded and used to leak data from various sensitive sources with use of user inputs (implicit permission). & \CheckmarkBold \\ \cline{2-4}
                                        & 13- Explicit  & The hub id (explicit permission) and state variables are leaked to an hard-coded phone number. & \CheckmarkBold \\ \cline{2-4}
                                        & 14- Explicit-Implicit  & The contact information (\ie phone number) is hard-coded to leak device information (implicit permission) and hub id (explicit permission).  & \CheckmarkBold \\ \cline{2-4}
                                         \hline
\multirow{3}{*}{\textbf{Multiple Leakage}}& 15- Multiple Leakage 1 & Various sensitive data obtained from state of the devices and user inputs and they are leaked via same sink interface. & \CheckmarkBold \\ \cline{2-4}
                                            & 16- Multiple Leakage 2 & Various sensitive data obtained from state of the devices and user inputs, and they are leaked via Internet and messaging sinks. & \CheckmarkBold \\ \cline{2-4} 
                                            & 17- Multiple Leakage 3 & Various sensitive obtained from state variables, and devices and they are leaked via more than one hard-coded contact information. & \CheckmarkBold\\ \cline{2-4} \hline 
\multirow{2}{*}{\textbf{Side Channel}}& 18- Side Channel 1 & A device operating in a  specific pattern is causing information leakage (e.g., on/off pattern of smart light). & \text{\sffamily !} \\ \cline{2-4}
                                         & 19- Side Channel 2 & A device operating in a specific pattern is causing another connected device to trigger some malicious activities. & \text{\sffamily !} \\ \hline\hline
\end{tabular}
\caption{Description of \iotbench test suite apps and \SainT's results.}
\label{appendix:iotbench-table}
\begin{tablenotes}
    \item[\textdaggerdbl] 19 apps leaks 27 sensitive data. We provide a comment block in the source code of the apps that gives detailed description of the leaks including the line number of the leaks and the ground truths.
    \vspace{2pt}
    \item[\textdagger]  \CheckmarkBold = True Positive, \text{\sffamily X} = False Positive, \text{\sffamily O} = Dynamic analysis required,  \text{\sffamily !} = Not considered in attacker model
\end{tablenotes}
\end{threeparttable}
\end{minipage}
\end{table*}

\end{document}